%% file: main.tex
\documentclass[9.5pt,journal,compsoc]{IEEEtran}
\IEEEoverridecommandlockouts
\usepackage{array}
\usepackage{gensymb}
\usepackage{booktabs} 
\usepackage{hyperref}
\usepackage{mathtools}
\usepackage{amsmath}
\usepackage{amssymb}
\usepackage{subcaption}
\usepackage{optidef}
\usepackage{bm}
\usepackage{amsfonts}
\usepackage{siunitx}
\usepackage{xcolor}
\usepackage{layouts}
\usepackage{subcaption}
\usepackage{caption}
\usepackage{multirow}
\usepackage{multicol}
\usepackage{graphicx}
\usepackage{array}
\usepackage[noend]{algpseudocode}
\usepackage{stmaryrd}
\usepackage[ruled,vlined,linesnumbered]{algorithm2e}

\def\BibTeX{{\rm B\kern-.05em{\sc i\kern-.025em b}\kern-.08em
    T\kern-.1667em\lower.7ex\hbox{E}\kern-.125emX}}

\begin{document}
\author{Sara~Garcia~Sanchez, Guillem~Reus~Muns, Carlos~Bocanegra,~Yanyu~Li,~Ufuk~Muncuk,~Yousof~Naderi, ~Yanzhi~Wang,~Stratis~Ioannidis~
and~Kaushik~R.~Chowdhury,~\IEEEmembership{Senior Member,~IEEE}
\thanks{S. Garcia Sanchez, G. Reus-Muns,  C. Bocanegra, Y. Li,  U. Muncuk, Y. Naderi, Y. Wang, S. Ioannidis and K. Chowdhury are with the Department of Electrical and Computer Engineering, Northeastern University, Boston,
MA, 02115 USA e-mail: (\{sgarcia, greusmuns, bocanegrac\}@coe.neu.edu, \{li.yanyu, u.muncuk\}@northeastern.edu, naderi@coe.neu.edu, yanz.wang@northeastern.edu, ioannidis@ece.neu.edu, krc@coe.neu.edu).}}

\title{AirNN: Neural Networks with Over-the-Air Convolution via Reconfigurable Intelligent Surfaces  }

\IEEEtitleabstractindextext{
\begin{abstract}
Over-the-air analog computation allows offloading computation to the wireless environment through carefully constructed transmitted signals. In this paper, we design and implement the first-of-its-kind over-the-air convolution and demonstrate it for inference tasks in a convolutional neural network (CNN). We engineer the ambient wireless propagation environment through reconfigurable intelligent surfaces (RIS) to design such an architecture, which we call 'AirNN'.  AirNN leverages the physics of wave reflection to represent a digital convolution, an essential part of a CNN architecture, in the analog domain. In contrast to classical communication, where the receiver must \textit{react} to the channel-induced transformation, generally represented as finite impulse response (FIR) filter, AirNN \textit{proactively} creates the signal reflections to emulate specific FIR filters through RIS. AirNN involves two steps: first, the weights of the neurons in the CNN are drawn from a finite set of channel impulse responses (CIR) that correspond to realizable FIR filters. Second, each CIR is engineered through RIS, and reflected signals combine at the receiver to determine the output of the convolution. This paper presents a proof-of-concept of AirNN by experimentally demonstrating  over-the-air convolutions. We then validate the entire resulting CNN model accuracy via simulations for an example task of modulation classification.
\end{abstract}

}

\maketitle
\input{sections/introduction-revised}
\input{sections/system_description}

\input{sections/AirNNconcept}

\input{sections/CNN}
\input{sections/IRS_simulator.tex}
\input{sections/setup}
\input{sections/performanceEval}
\input{sections/conclusions}

\bibliographystyle{ieeetr}


\vspace{-10mm}
\begin{IEEEbiography}[{\includegraphics[width=1.1in,height=1.35in,clip,keepaspectratio]{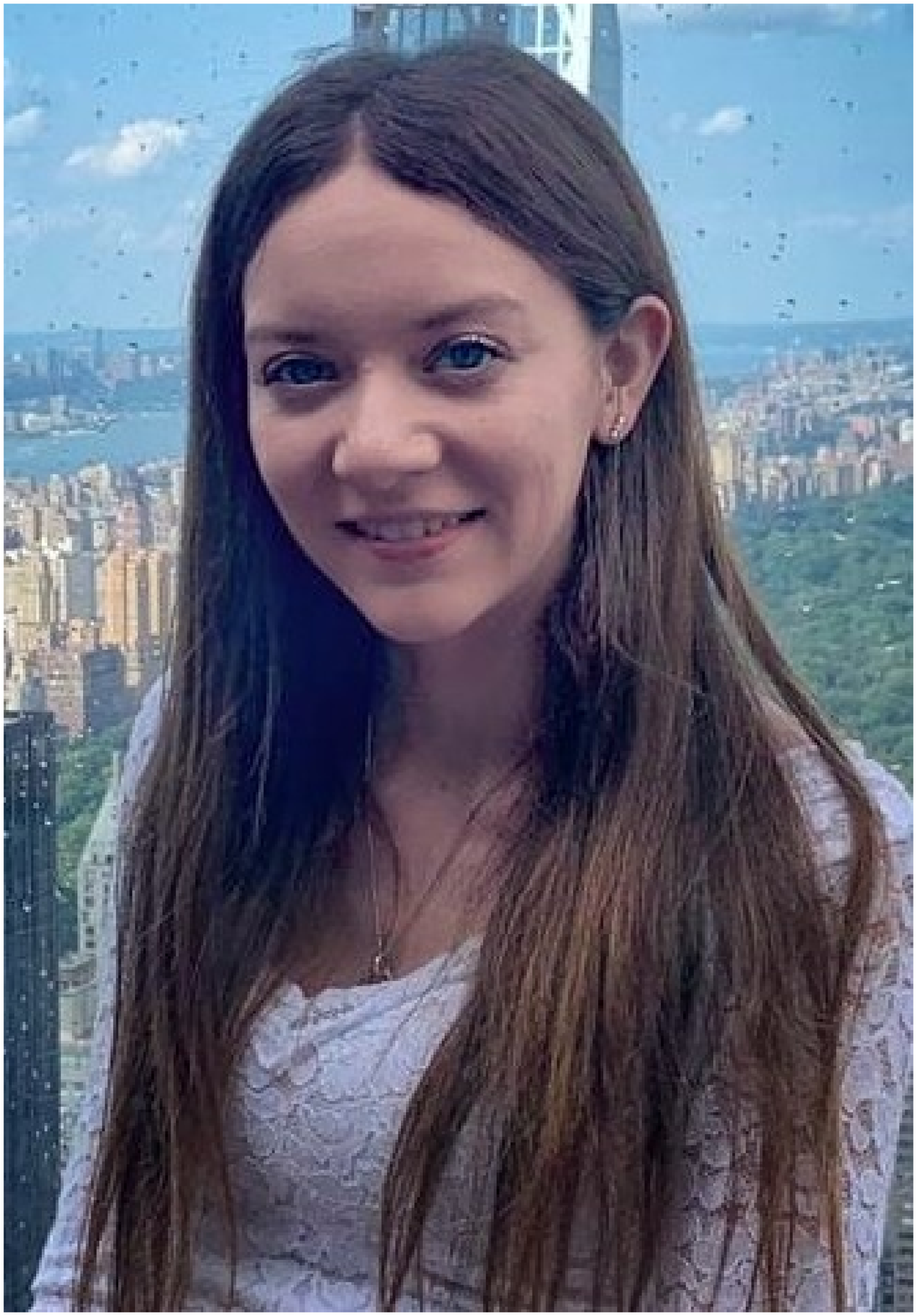}}]{Sara Garcia Sanchez} received the B.S. and M.S. degrees in Electrical Engineering from Universidad Politecnica de Madrid in 2016 and 2018 respectively. She is currently a PhD candidate at the Department of Electrical and Computer Engineering in Northeastern University, under the guidance of Professor Kaushik Roy Chowdhury. Her research interests include mmWave, UAV communications, MIMO and optimization techniques.
\vspace{-20mm}
\end{IEEEbiography}

\begin{IEEEbiography}[{\includegraphics[width=1.1in,height=1.35in,clip,keepaspectratio]{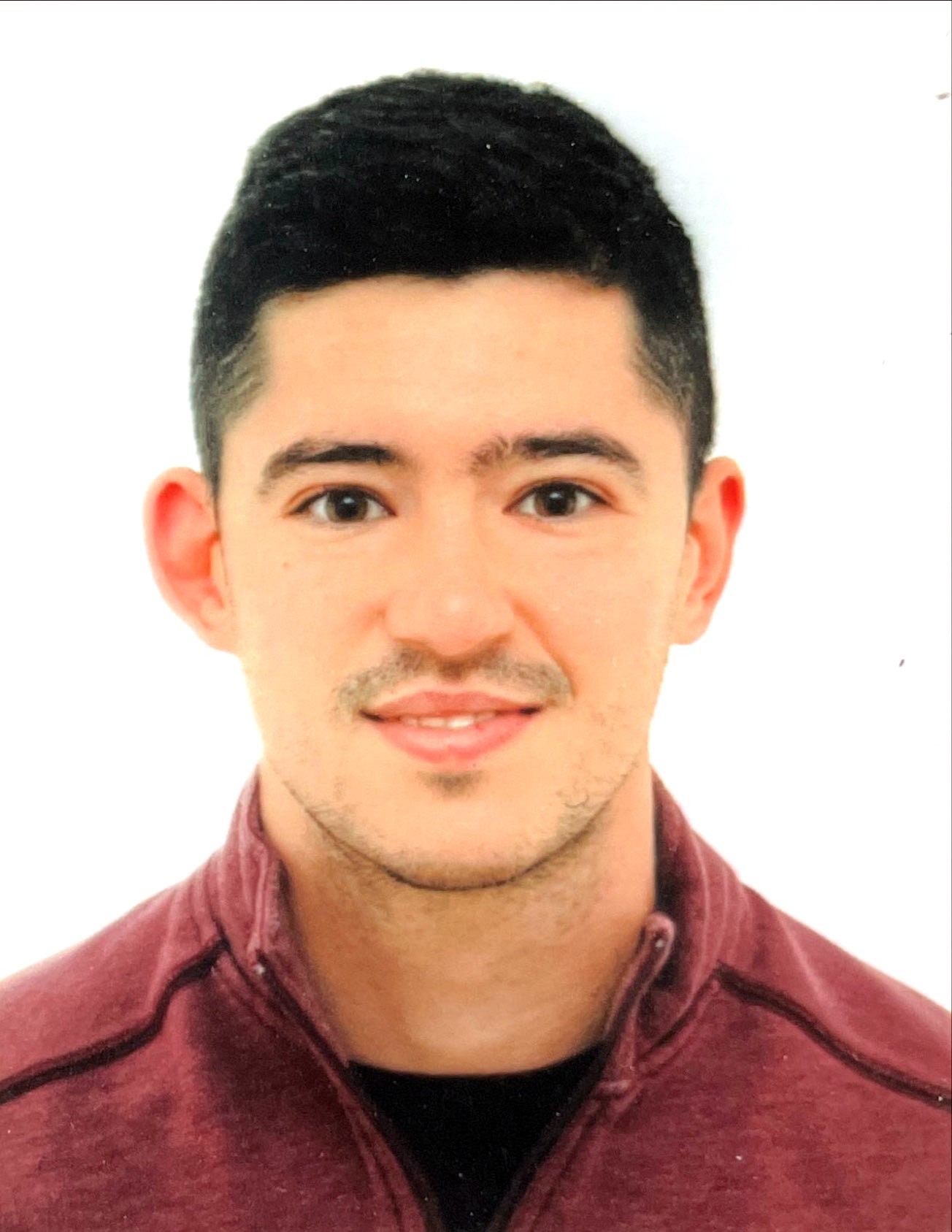}}]{Guillem Reus Muns} received the B.Sc. degree in telecommunications engineering from the Polytechnic University of Catalonia (UPC-BarcelonaTech). He joined Northeastern University, USA, in 2018, where he got his M.Sc. in electrical and computer engineering and is currently working towards his Ph.D. His current interests include mobile communications, networked robotics, machine learning for wireless communications and spectrum access.
\vspace{-20mm}
\end{IEEEbiography}

\begin{IEEEbiography}[{\includegraphics[width=1.1in,height=1.35in,clip,keepaspectratio]{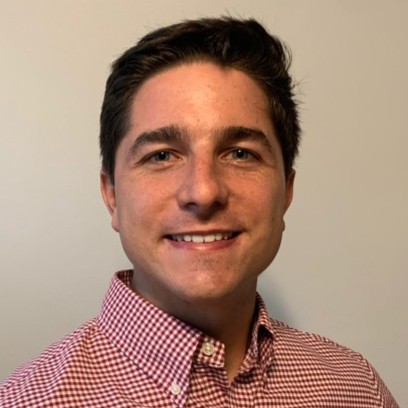}}]{Carlos Bocanegra} received the B.S. degree in Telecommunications Engineering from the Polytechnic University of Catalonia (UPC), Barcelona Spain, in 2015, M.S. in Electrical Engineering from Northeastern University, Boston MA, in 2017, and Ph.D. in Computer Engineering from Northeastern University, in 2021. Currently, he holds a position of Senior Wireless System Engineer under Delart Tech working at Facebook Reality Labs (FRL) where he devises simulation frameworks to study coexistence, latency, and power consumption for 5G NR and IoT. He has also conducted research and product development in companies such as NEC Laboratories America, Princeton, NJ; and MathWorks, Natick, MA. His research interests include the design and prototype of multi-antenna systems, coexistence in HetNets, machine learning for wireless applications, and virtualization at the RAN using Software Defined Radios (SDR).
\vspace{-20mm}
\end{IEEEbiography}

\begin{IEEEbiography}[{\includegraphics[width=1.1in,height=1.35in,clip,keepaspectratio]{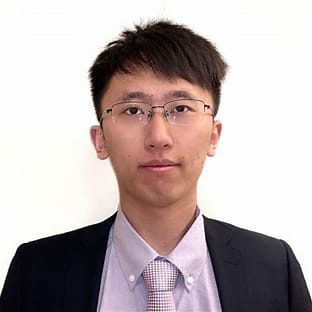}}]{Yanyu Li} is a Ph.D. candidate at the Department of Electrical and Computer Engineering in Northeastern University, advised by Professor Yanzhi Wang. His research interests include deep learning, neural network architecture search, pruning and quantization.
\vspace{-12mm}
\end{IEEEbiography}
\begin{IEEEbiography}[{\includegraphics[width=1.1in,height=1.35in,clip,keepaspectratio]{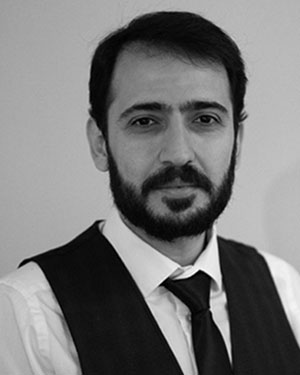}}]{Ufuk Muncuk} received the Ph.D. degree in electrical and computer engineering from Northeastern University, Boston, MA, USA, in 2019. He is a Research Assistant Professor with the Electrical and Computer Engineering Department, Northeastern University. His research interests include design, optimization, and implementation for RF energy harvesting circuits and system design for RF energy and magnetic coupling-based energy transfer, intrabody transceivers, and cognitive radio systems. Dr. Muncuk was recipient of the Best Paper Runners-Up at ACM SenSyS in 2018, and Best Paper Awards at IEEE ICC in 2013 and IEEE GLOBECOM in 2019.  
\vspace{-15mm}
\end{IEEEbiography}

\begin{IEEEbiography}[{\includegraphics[width=1.1in,height=1.35in,clip,keepaspectratio]{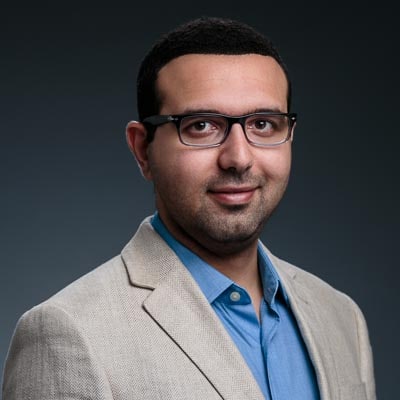}}]{Yousof Naderi} is a Research Assistant Professor in the Electrical and Computer Engineering Depart- ment at Northeastern University, Boston, MA. He received the Ph.D. degree in Electrical and Computer Engineering from Northeastern University, Boston in 2015. He was the recipient of NEU Ph.D. dissertation award in 2015, a finalist in the Bell Labs Prize competition in 2017, Best Paper Awards at the IEEE INFOCOM in 2018 and IEEE GLOBECOM in 2019. His research expertise lies in the design and development of AI-powered cyber-physical systems, intelligent surfaces for 6G and beyond, and self-powered networked robotics. 
\vspace{-10mm}
\end{IEEEbiography}
 \begin{IEEEbiography} [{\includegraphics[width=1.1in,height=1.35in,clip,keepaspectratio]{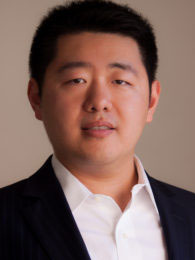}}]
{Yanzhi Wang} is currently an Assistant Professor at the Department of ECE at Northeastern University, Boston, MA. His research focuses on model compression and platform-specific acceleration of deep learning architectures, maintaining the highest model compression rates on representative DNNs since 09/2018. His work on AQFP superconducting based DNN acceleration is by far the highest energy efficiency among all hardware devices. His recent research achievement, CoCoPIE, can achieve real-time performance on almost all deep learning applications using off-the-shelf mobile devices, outperforming competing frameworks by up to 180X acceleration.
He received the U.S. Army Young Investigator Program Award (YIP), Massachusetts Acorn Innovation Award, Ming Hsieh Scholar Award, and other research awards from Google, MathWorks. etc.
\vspace{-15mm}
\end{IEEEbiography}
\begin{IEEEbiography}[{\includegraphics[width=1.1in,height=1.35in,clip,keepaspectratio]{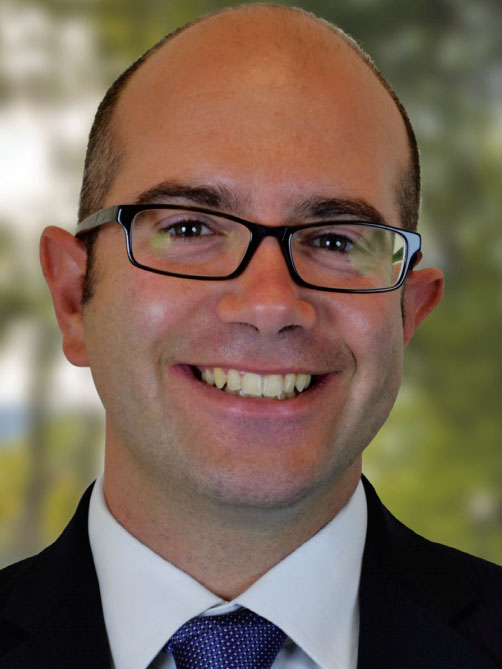}}]{Stratis Ioannidis} is an Associate Professor in the Electrical and Computer Engineering Department of Northeastern University, in Boston, MA, where he also holds a courtesy appointment with the Khoury College of Computer Sciences.
Prior to joining Northeastern, he was a research scientist at the Technicolor research centers in Paris, France, and Palo Alto, CA, as well as at Yahoo Labs in Sunnyvale, CA.
His research interests span machine learning, distributed systems, networking, optimization, and privacy.
\vspace{-10mm}
\end{IEEEbiography}
\begin{IEEEbiography}[{\includegraphics[width=1.1in,height=1.35in,clip,keepaspectratio]{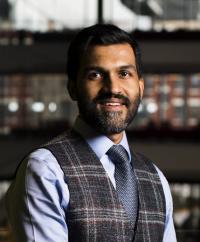}}]{Kaushik Roy Chowdhury} is a Professor at Northeastern University, Boston, MA.
He is presently a co-director of the Platforms for Advanced Wireless Research (PAWR) project office. His current research interests involve systems aspects of networked robotics, machine learning for agile spectrum sensing/access, wireless energy transfer, and large-scale experimental deployment of emerging wireless technologies.
\end{IEEEbiography}
\end{document}

%% file: sections/introduction-revised.tex
\section{Introduction}
\label{sec:introduction}

New and emerging Internet of Things (IoT) applications require collecting and processing large amounts of data, generally transmitted over the wireless channel. In this context, \textit{over-the-air} analog computation has been proposed as a alternative to all-digital approaches using acoustic~\cite{hughes2019wave}, optical~\cite{sui2020review} and RF~\cite{elbir2020survey} signals. 
The core idea is to take advantage of additional degrees of freedom in the environment to partially offload computation into the wireless domain. Ideally, communications signals that carry required information from the source are also controlled and modified, such that the received signal emulates the end result of a mathematical operation. Recent results, albeit limited to pure simulation studies, have demonstrated remarkable promise for operations like data aggregation \cite{yu2020optimizing} and processing in recurrent neural networks \cite{hughes2019wave}.

\begin{figure}[t]
\centering
\begin{subfigure}{\linewidth}
  \centering
  \includegraphics[trim=75 200 250 15,clip,width=\linewidth]{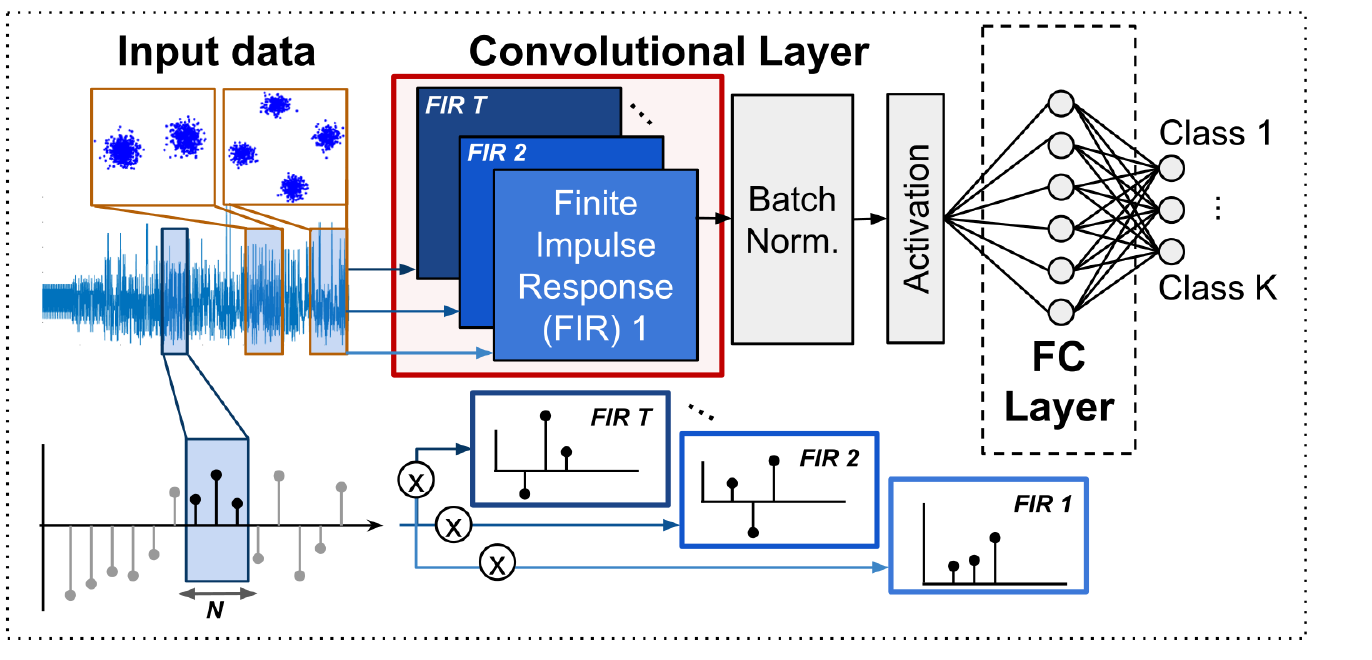}
  \caption{}
  \label{fig:intro1}%
\end{subfigure}
\begin{subfigure}{\linewidth}
  \centering
  \includegraphics[trim=75 10 250 190,clip,width=\linewidth]{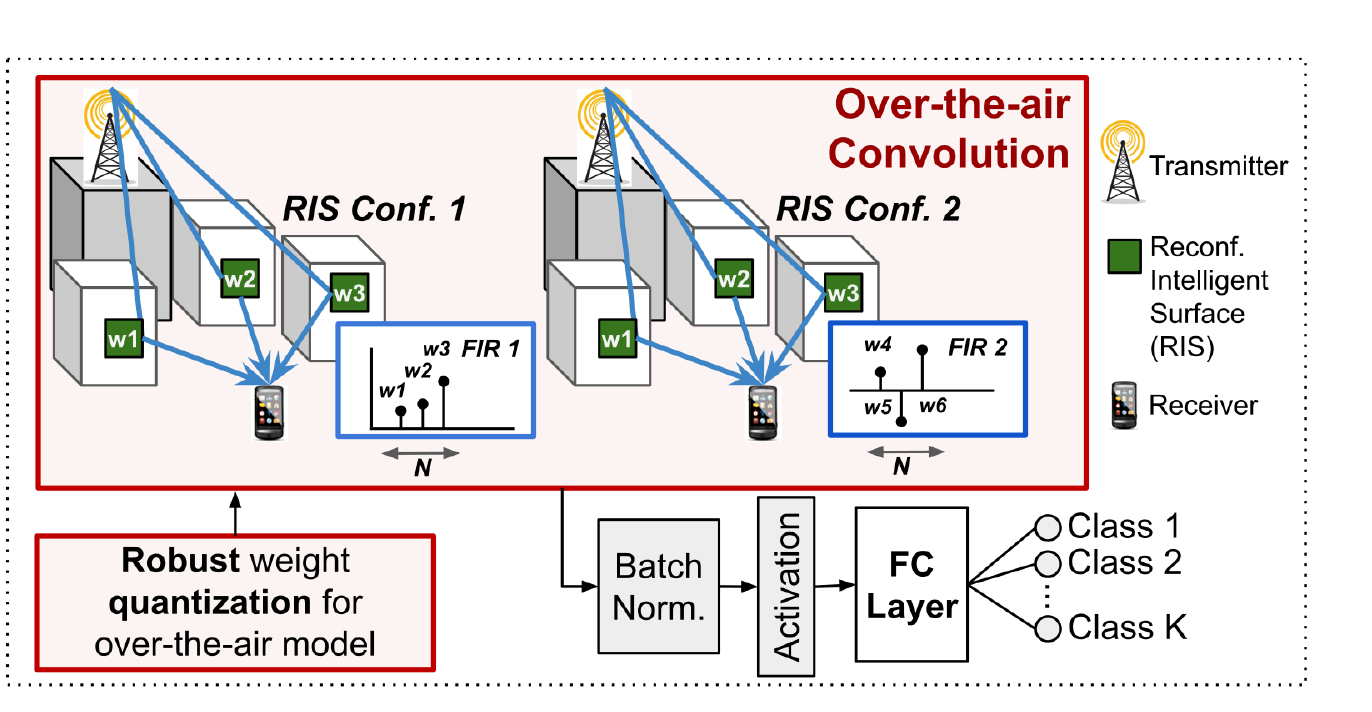}
  \caption{}
  \label{fig:intro2}%
\end{subfigure}
\caption{\small (a) Conventional CNN architecture, highlighting the convolution step, with input data in the form of raw IQ samples and digital convolution operations in software that are represented as a bank of FIR filters (shown in red box), (b) AirNN architecture shows the same convolution operation but over-the-air, using a RIS network. Different RIS configurations result in specific channel transformations equivalent to the convolution operation shown in (a).}
\label{intro}
\end{figure}

The wireless research community has applied machine learning (ML) methods for physical layer related problems of protocol classification~\cite{zhao2017waveforms}, adversarial activity detection, modulation classification~\cite{cai2019modulation} and RF fingerprinting~\cite{restuccia2019deepradioid}, among others. In particular, the ML solutions proposed in these works are based on a special class of architectures called as convolutional neural networks (CNNs). Fig.~\ref{fig:intro1} shows such an example processing chain, where raw in-phase/quadrature (IQ) samples are fed to the neural network composed of a convolutional layer, followed by a fully connected (FC) layer that predicts the signal modulation type.

Given the interest in applying CNNs on RF signals and the promise of analog computation, this paper poses the following question: \textit{what if we were able to realize analog over-the-air convolutions accurately enough to substitute its digital equivalent in a CNN?} We describe a methodology to achieve this objective and demonstrate it experimentally. We then show how this over-the-air operation impacts more complex mathematical computations, such as a CNN (that may have hundreds of such convolution operations). 
 
\noindent $\bullet$ \textbf{Programming the Environment:}
We propose a radically different approach by shifting the burden of executing the convolution operation from dedicated digital devices into the ambient environment: several copies of a transmitted wireless signal interact with a carefully engineered propagation and reflection environment to emulate the mathematically equivalent outcome of passing the signal through a digital \textit{convolution filter} present in a CNN. Each copy is modified through a reflection using a reconfigurable intelligent surface (RIS)~\cite{dunna2020scattermimo}. Once all these copies combine, the cumulative effect at the receiver resembles the processing of the same input signal as if it passed through a convolutional layer  used in a CNN. As this step happens over-the-air, we refer to the resulting architecture as 'AirNN'.  We can extend this concept from a \textit{single} convolution computation to a number of them performed in succession. Fig.~\ref{fig:intro2} shows two configurations of RIS that give rise to two different desired channel impulse responses (CIRs), each of which acts as an individual finite impulse response (FIR) filter convolving with the transmitted signal. The filter representation can be made to exactly match the filters learned during training a classical, digital CNN.

\noindent $\bullet$ \textbf{Challenges in Designing AirNN:}
While the domain of analog computation has existed for over a decade~\cite{boser1991analog}, combining wireless signals to emulate a digital convolution operation has not been attempted before. It is noted that AirNN relies on representing a convolutional filter of size $N$ in a CNN as an \textit{N}-tap FIR filter. This leverages the mathematical equivalence between the latter and the $N$ tap discrete version of the CIR. In order to realize this equivalence in practice, we identify several challenges that need to be addressed. First, CIR depends on the transmitted signal and the multipath components of the environment, which the RIS can influence to a significant extent, but not perfectly. This motivates the design of an efficient optimization loop: we must be able to train a CNN with 
quantized weights, drawn from a very limited candidate set, that correspond to the feasible CIR set that can be attained trough the use of RIS in practice. This mapping between RIS configuration and CIR deviates over time as the wireless channel conditions change. Therefore, we need to engineer repeatable conditions during testing while accommodating ambient factors that cannot be controlled. Second, from a systems viewpoint, we need to create a network of programmable, low-cost RIS that is time-synchronized and responds to control directives to change its reflection ability. Finally, a CNN with experimentally computed convolution in AirNN should demonstrate accuracy comparable to its all-digital CNN running on a GPU. 

\noindent $\bullet$ \textbf{Summary of Contributions in AirNN:}
Our main contributions are as follows:

\noindent (1) {We formulate and experimentally demonstrate the theory that maps digital (processing-based) and analog (over-the-air) convolutions using programmable RIS. }

\noindent (2) {We propose a method to train CNNs with a quantized set of weights drawn from the RIS-generated candidate set without appreciable loss of accuracy for a task of modulation classification, compared to unconstrained training. We include measures to increase resiliency when the wireless channel changes over time.}

\noindent (3) As a systems contribution, we implement a software-framework to control the RIS network called AirNNOS that synchronizes and aligns start times of the relay transmitters and the receiver, as well as reconfigures the RIS on demand to change their reflection coefficients. 

\noindent (4) Given the measured error of the over-the-air convolution, we show through simulations that the experimentally derived analog convolution is accurate enough to run inference on trained neural networks, with an average deviation in testing accuracy of 3.2\%  for a range of medium-to-high SNR of [6, 30] dB 
compared to classical, GPU-based inference.

%% file: sections/system_description.tex
\section{Related Work}
The area of analog computing is in a nascent stage. A variety of approaches spanning digital, analog, hybrid and FPGA-based solutions have been studied to accelerate training and inference in NN \cite{misra2010artificial}. The authors in \cite{kendall2020training} propose a method to train end-to-end analog NN using stochastic gradient descent by varying the conductance of programmable resistive devices and diodes. In-situ learning for a memristor-based multi-layer perceptron is demonstrated in \cite{li2018efficient}. The authors in \cite{hughes2019wave} leverage wave physics properties to experimentally demonstrate an analog recurrent NN using acoustic signals. The neural network-based analog computing has been utilized in various 
other applications such as
super-resolution memristor crossbar \cite{james2021analog},
Charge-Trap-Transistor (CTT) computing engine \cite{du2018analog}, 
scalable VLSI \cite{cauwenberghs1996analog}, FPGA-based extreme-throughput operations \cite{umuroglu2020logicnets}, and neuromorphic engineering \cite{strukov2019building}.

Specific to the RF domain, Over-the-Air Computation (AirComp) has been receiving special attention for solving problems such as data aggregation \cite{wen2019reduced}, efficient battery recharging trough beamforming \cite{wang2020wirelessly}\cite{arun2020rfocus} and local model uploading for federated learning \cite{liu2021joint} \cite{ni2020intelligent}. Several of these works propose as part of their solution the use of RIS to enhance channel conditions by compensating for destructive interference \cite{wang2021federated}, enhancing the received power \cite{jiang2019over} or  maximizing the achievable hybrid rate of all users in a network  \cite{ni2021over}. However, all above works are validated in simulation only and thus, they do not provide any insight on the implementation feasibility of their approaches. 

Different from these works, that mostly rely on AirComp to perform data aggregation tasks, this paper is, to the best of our knowledge,
the first demonstrator of using RF signals and RIS to replicate  digital convolutions over-the-air with validation on a prototype testbed. Our focus on the convolution operation is motivated by the remarkable performance that CNNs have shown within the deep learning community in fields such as computer vision, signal processing or RF signal classification \cite{lecun2015deep}. This performance has attracted numerous research efforts towards realizing convolution implementations, as this processing step alone consumes over 80\% of the total computation during the forward propagation step \cite{li2016performance}

\section{AirNN Operational Overview}  \label{sec:airnn_system_overview}

In AirNN, we perform an over-the-air convolution making use of a network of programmable RIS and a relay node, as shown in Fig.~\ref{fig:systemArch}. The relay node receives the incoming signal from the Tx (shown by link \texttt{A}). Following this, the relay forwards that same signal using directional antennas \texttt{R-Tx} (shown by link \texttt{B}) towards specific RIS. 
The network controller (\texttt{NNCtrl}) adjusts the reflection angles of the different RIS to ensure that the reflected signals (shown by link \texttt{C}) combine in a deterministic manner at the receiver antenna \texttt{R-Rx} of the relay. 
Thus, the signals shaped by the RIS emulate the outcome of a digital convolution at the receiver antenna \texttt{R-Rx} of the relay. The \texttt{NNCtrl} block is tasked to train the network with a quantized set of weights, dictated by the set of reflections that our network of RIS can generate. 
During inference, the \texttt{NNCtrl} notifies the RIS network with the updated RIS configuration, i.e., that results in the desired convolution, using a dedicated control plane that connects the relay to the microcontrollers (see Fig.\ref{fig:systemArch}).\\

%% file: sections/AirNNconcept.tex
\section{Over-the-Air Convolution Theory}
\label{sec:relationship}

This section explains the theory behind AirNN, namely,  how to map digital convolution to over-the-air signal transformation by the wireless channel. 

\subsection{Mapping the Process of Convolution} \label{subsec:convolutionSimilarities}

\noindent$\bullet$ \textbf{Convolution in a Digitally Constructed CNN:} In a given CNN, such as the example shown in Fig.~\ref{fig:intro1}, the convolutional filters are learned during the training process. These filters activate the neurons when a specific feature of interest is detected during testing. For 1-D inputs to the CNN, typical for streaming IQ samples from a wireless signal, these filters can be represented as an FIR of length $N$ i.e. $N$ taps, filter order $L = N-1$. This filter is essentially a vector of $N$ complex weights, each weight defining a specific amplitude and phase of that particular filter tap. As an example, consider the output of a filter of length $N$ 
    \label{eq:FIR}
in Eq.~\ref{eq:FIR},
 where $\textbf{w}=\{w_0,w_1,...,w_{N-1}\} \in \mathbb{C}$ are  the complex weights that are applied to the incoming stream of samples.
The filter order $L$ also gives the number of input samples needed to generate a single sample at the output.
\begin{equation}
    y[n] = w_0*x[n+\frac{L}{2}] + \cdots + w_{\frac{L}{2}}*x[n] + \cdots + w_{L}*x[n-\frac{L}{2}],
    \label{eq:FIR}
\end{equation}

\noindent$\bullet$ \textbf{Convolution in the Wireless Channel:}
Our goal is simple: we wish to artificially construct a signal transformation in the physical environment during testing that precisely maps to the above vector $\textbf{w}$ that we obtained during training time. We leverage the fact that when a signal is transmitted over the air, the reflections from the environment cause copies of the same signal to arrive at the receiver with different amplitude, phase, and time delays, collectively referred to as \textit{multipath}. This phenomenon is characterized by the CIR, where each path is defined by the tuple of complex transformations in amplitude and phase and the instant of arrival at the receiver. This multipath results in an FIR filter of order $N-1$, where $N$ is the total number of paths. 
 Here, the first path is associated with the Line of Sight (LoS) component, whereas the $N-1$ later paths arise from Non-Line of Sight (NLoS). 

 \begin{figure}[t]
    \centering
    \includegraphics[trim=100 70 50 50,clip,width=\linewidth]{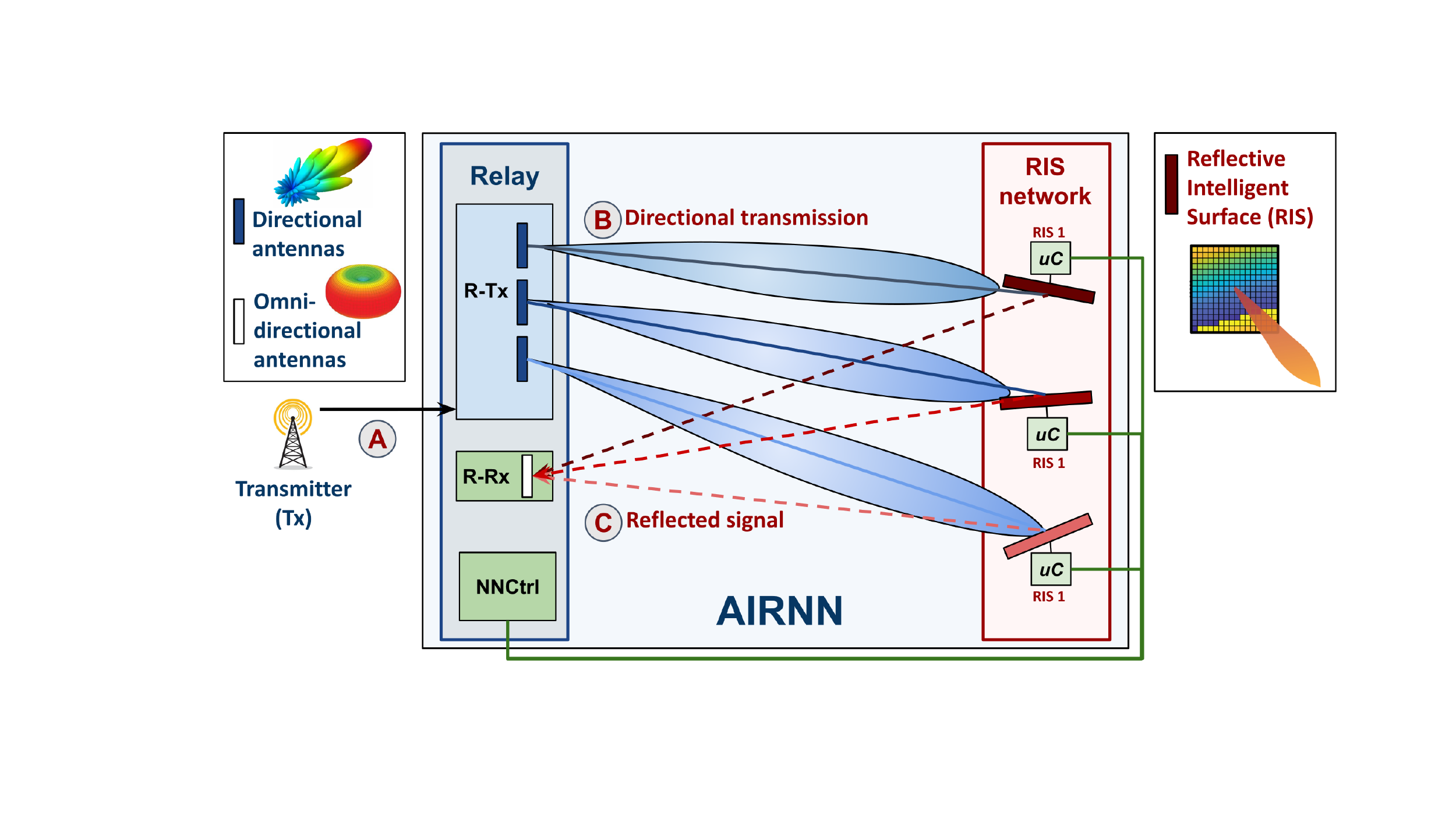}
    \caption{\small AirNN system components and transmission sequence: 
    external device Tx transmits the signal of interest. The relay node captures this signal and uses antennas (\texttt{R-Tx})  to forward it to the network of programmable RIS controlled via the neural network controller (\texttt{NNCtrl}) module. The RIS reflect the signal with desired channel transformations to the relay's antennas (\texttt{R-Rx}). 
    }
    \centering
\label{fig:systemArch}
\end{figure}

\subsection{Engineering Convolutions using RIS}

Each RIS is a planar array of passive reflective antennas, and each such antenna has a selectable range of impedance matching circuits. These circuits are programmable, and by activating one over the others, we change the impedance of the corresponding reflective antenna. This alters the antenna reflection coefficient, which then changes the phase of the reflected signal.
The RIS-guided reflections allow flexibility in imparting the desired complex-valued amplitude and phase changes to the signal. However, the set of candidate options is limited, i.e., the feasible code-book is constrained by the number of available RIS, the selectable circuit combinations within each RIS reflective antenna array, and the geometry of the propagation environment.

\subsection{Systems Challenges in AirNN} 
\label{subsec:challenges}
While the concept of AirNN is intuitive, there are several systems challenges for practical realization, as we briefly covered in the introduction and further describe below. 

\noindent \textbf{(Ch1) Complex-Valued Convolutions}: Complex numbers are used jointly to represent amplitude and phase information in the RF domain. Thus, mapping real-valued convolutional layer filters to the complex-valued CIR is not feasible. We can only use complex-valued neural networks, as we describe in Sec.~\ref{subsec:complexNN}.

\noindent \textbf{(Ch2) RIS-Based Weight Constraints}: The number of possible FIRs that we can engineer via RIS is limited. In the digital domain, this constrains the set of feasible FIR filters that can be used during the training of the CNN. Thus, AirNN must quantize the CNN weights that correspond to only possible (i.e., RIS-realizable) FIR filters, as given in Sec.~\ref{subsec:constrainedCNN}.

\noindent \textbf{(Ch3) Receiver Noise}: Even if the channel remains time-invariant and the RIS configuration is static, there exists thermal noise. We need to account for this stochastic noise, especially as the reflected signals are low in amplitude and barely above the noise floor. We explain how we achieve this for additive white Gaussian noise via a correction factor in Sec.~\ref{subsec:dataAug}.

\noindent \textbf{(Ch4) RIS-Path Separation}: The FIR filter taps that we obtain through AirNN must be equally spaced in time, as is also assumed in the digital version. In the wireless domain, this is challenging as the arrival time of the signal depends on separation distances and the sampling rate. AirNN addresses this via a multi-antenna relay (see Fig.~\ref{fig:systemArch},  where \texttt{R-Tx} has three elements) that ensures sufficient path separation. We explain this in Sec.~\ref{subsec:multi_antenna}.

\noindent \textbf{(Ch5) Meaningful CIR Variations}: The LoS path dominates over the NLoS paths resulting from RIS reflections in terms of received signal strength.  To ensure that the artificially constructed NLoS paths shape the CIR precisely (despite the overbearing LoS path), we use directional antennas at the relay \texttt{R-Tx} as explained in Sec.~\ref{subsec:direct_antenna}.

\noindent \textbf{(Ch6) Channel Variations}: If the wireless channel changes, then the prior configured RIS may still generate an older and outdated CIR.  To prevent both re-training the neural network or re-configuring the RIS, AirNN compensates for channel variations from a pre-determined baseline, as we show in Sec.~\ref{CIR_offset}.

\noindent \textbf{(Ch7) Precise Synchronization}: Given the concise time window to achieve convolution, each  antenna must adjust the start times to achieve $\mu$s-level synchronization for Mbps-level data rate. Long symbol times can disrupt the system as the CIR may change beyond the estimated value. AirNN solves this problem by padding the sequence at the \texttt{R-Tx} with zeroes to achieve precisely one sample delay between any two successive signals, as detailed in Sec.~\ref{subsec:airnnnos}.

%% file: sections/CNN.tex
\section{AirNN Neural Network Design}
\label{sec:contrained_CNN}
This section explains how we design AirNN by addressing the design challenges Ch1, Ch2 and Ch3. Remaining challenges are addressed in Sec.~\ref{Simu_Exp}.
\subsection{Design Complex-Valued CNN (Ch1)}
\label{subsec:complexNN}
To facilitate the mapping between the neural network weights and the RIS-engineered CIR, we design a neural network model based on complex-valued data and weights \cite{trabelsi2018deep}. 
Given that the convolution operator ($*$) is distributive, the output of a complex convolutional layer $\phi$ can be expressed as: 
\begin{equation}
 y = \phi_{w_R}(x_R) - \phi_{w_I}(x_I) + j(\phi_{w_I}(x_R)+\phi_{w_R}(x_I))
\end{equation}
where $y$ is the output of the complex convolution, $x$ and $w$ represent the input and weights of the convolutional layer and $x_{R/I}$, $w_{R/I}$ are the real/imaginary parts of $x$ and $w$, respectively. The distributive property also applies to the product-sum operation of fully connected (FC) layers. Thus, we design complex-valued layers ($\phi_w$) using two real-valued layers, where each one of them will independently represent the real ($\phi_{w_R}$) and imaginary parts ($\phi_{w_I}$). 
The seminal work in \cite{trabelsi2018deep} provides a detailed explanation of complex neural network theory and implementation.

\subsection{Constrained Weight Quantization (Ch2)}
\label{subsec:constrainedCNN}

We use a quantization-enabled approach to train the neural network with the set of feasible weights provided by the RIS-engineered environment. 
Let the weights of a complex convolution layer be:

\begin{equation}
\begin{split}
    W=&\{w_1, ...,  w_F\}, \quad w_f \in \mathbb{C}^{N} \\
    w_f =& [w_f^1, ..., w_f^{N}], \quad w_f^n \in \mathbb{C}, 
\end{split}
\end{equation}

with $w_f \in \mathbb{C}^{N}, w_f^n \in \mathbb{C}$ and where $W$ is the set of $F$ FIR filters ($w_f$) with length $N$ that represent the layer weights. As described in Sec.~\ref{subsec:convolutionSimilarities}, AirNN has limited freedom in implementing an over-the-air FIR filter. 
Therefore, we constrain the weights $w_f^n$ for each filter tap with index $n$ to a candidate set $S_n$ of implementable values, defined as: 

\begin{equation}
    S_n = \{ c_1^n, ..., c_s^n, ..., c_{|S_n|}^n \}, \quad c_{s}^n \in \mathbb{C}, \quad 1 < n< N 
\end{equation}

$|S_n|$ is the size of the constrained set and $c_s^n$ represents each of its complex-valued elements.
During training, we compute the Euclidean distance (D) from every individual weight $w_f^n \in W$ to all weight candidates $c_s^n \in S_n$. Then, we define the nearest neighbor of $w_f^n$ as:
\begin{equation}
    w_{f}^{n}{'} = \arg \min_{c \in S_n} D (c, w_{f}^{n}).
\label{eq:euclidean}
\end{equation}
While training the model, the weight values $w_f^n$ are rounded to their nearest neighbors $w_{f}^{n}{'}$ to perform forward propagation, following Eq.~\ref{eq:euclidean}. 
However, the derivative of the rounding function is zero throughout and cannot be trained via classic backpropagation. 
We solve this by employing the  Straight Through Estimator (STE) approach ~\cite{DBLP:journals/corr/BengioLC13, DBLP:journals/corr/abs-1903-05662}, which assumes the derivative of the discrete rounding function to be $\boldsymbol{1}$. While other approaches based on ADMM \cite{chang2020mix} have also been proposed, we select STE due to its faster training and convergence. Then, the forward and backward propagation steps can be expressed as:
\begin{equation}
\begin{aligned}
    \textbf{Forward:}\  &\mathcal{L}=\phi_{w'}(input);\quad \textbf{Backward:}\ \frac{\partial \mathcal{L}}{\partial w} = \frac{\partial \mathcal{L}}{\partial w'}
\end{aligned}
\label{eq:ste}
\vspace{-1mm}
\end{equation}
where $\mathcal{L}$ can be any form of loss function. 
Here, the gradient of $w$ is approximated to the gradient of $w'$, which is the fundamental working principle of STE.

\subsection{Handling Errors in Weights (Ch3)}
\label{subsec:dataAug}
As mentioned in Sec.~\ref{subsec:challenges}, the receiver introduces thermal noise that causes random variations, denoted henceforth as $\epsilon \in \mathbb{C}$, into the RIS-implemented CIR. 
Such CIR variations follow a Gaussian distribution with standard deviation $\sigma$, i.e., $\epsilon \sim \mathcal{CN}(0, \sigma^2)$ \cite{Gaussian}.

Due to noise and changing wireless environment, the current CIR may have a mismatch with the filters identified by the RIS, and yet we desire the CNN to be robust without appreciable fall in accuracy.
In order to solve this problem, we modify Eq.~\ref{eq:euclidean} by adding the term $\epsilon$, as given below:
\begin{equation}
    w_{f}^{n}{'} = \arg \min_{c \in S_n} D (c, w_f^n)
 + \epsilon.
\end{equation}
As opposed to previous data augmentation approaches, the variable $\epsilon$ is applied during training directly to the weights to increase the robustness of the model as well as during testing.
In each forward propagation step, weights are first quantized to the target constraint and noise is added. 
After the forward loss has been computed, we use backpropagation and obtain gradients for $w’$.
As mentioned previously, STE is employed to approximate gradients for $w$, such that $w$ is updated via Stochastic Gradient Descent (SGD).

%% file: sections/IRS_simulator.tex
\section{AirNN Relay Transmitter  Design}  \label{Simu_Exp}

In this section, we design the relay transmitter \texttt{R-Tx} and address the challenges Ch4, Ch5 and Ch6  described in Sec. \ref{subsec:challenges}.

\begin{figure}[t!]
\label{fig:simuIRS}
\centering
\begin{minipage}{.49\linewidth}
    \begin{minipage}{.49\linewidth}
        \begin{subfigure}{\linewidth}
            \includegraphics[width=\linewidth]{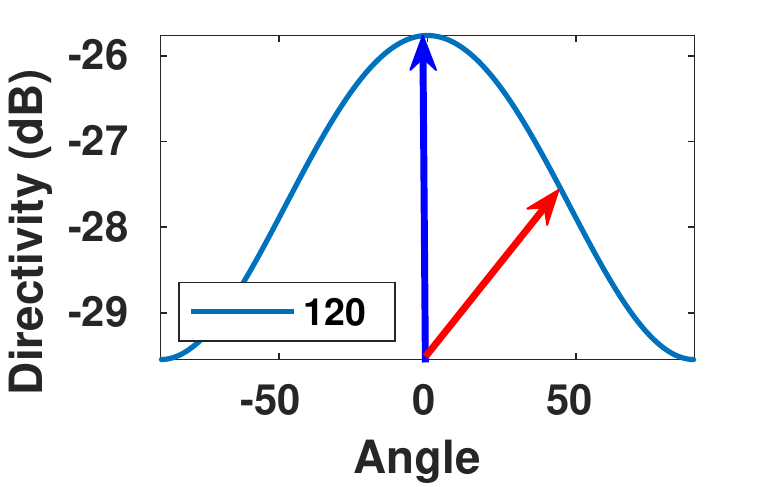}
            \label{subfig:soa_bin}
        \end{subfigure}
    \end{minipage}
    \begin{minipage}{.49\linewidth}
        \begin{subfigure}{\linewidth}
            \includegraphics[width=\linewidth]{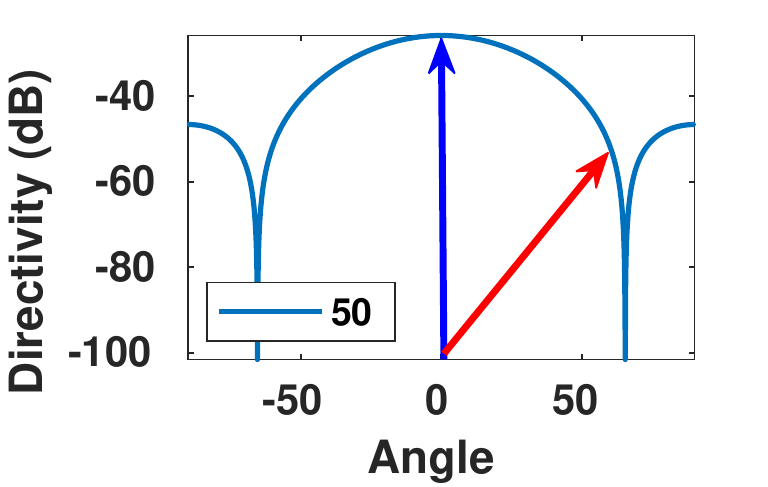}            \label{subfig:soa_surface}
        \end{subfigure}
    \end{minipage}
    \begin{minipage}{.49\linewidth}
        \begin{subfigure}{\linewidth}
            \includegraphics[width=\linewidth]{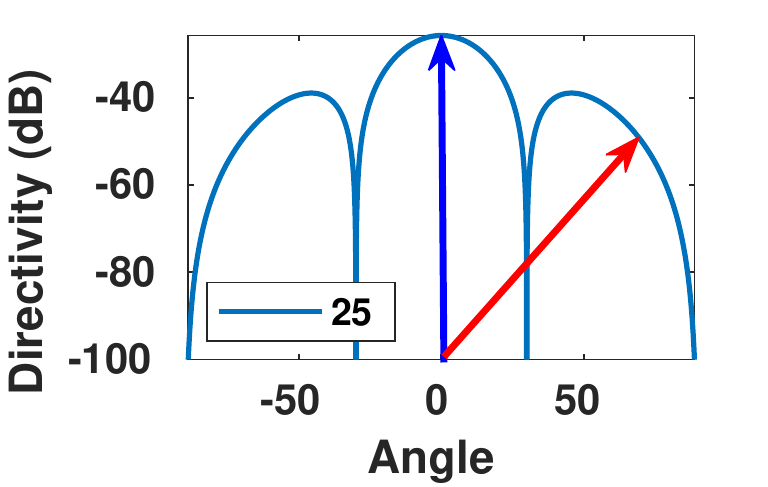}
            \label{subfig:soa_bin}
        \end{subfigure}
    \end{minipage}
    \begin{minipage}{.49\linewidth}
        \begin{subfigure}{\linewidth}
            \includegraphics[width=\linewidth]{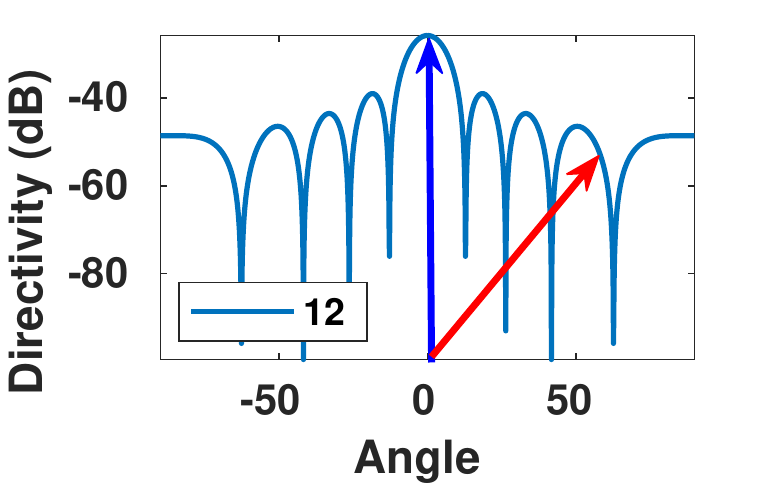}
            \label{subfig:direct}
        \end{subfigure}
    \end{minipage}
    \vspace{1.75mm}
    \subcaption{Directivities}
\end{minipage}
\begin{subfigure}{0.495\linewidth}
  \centering
     \includegraphics[trim=0 0 20 180,clip,width=\linewidth]{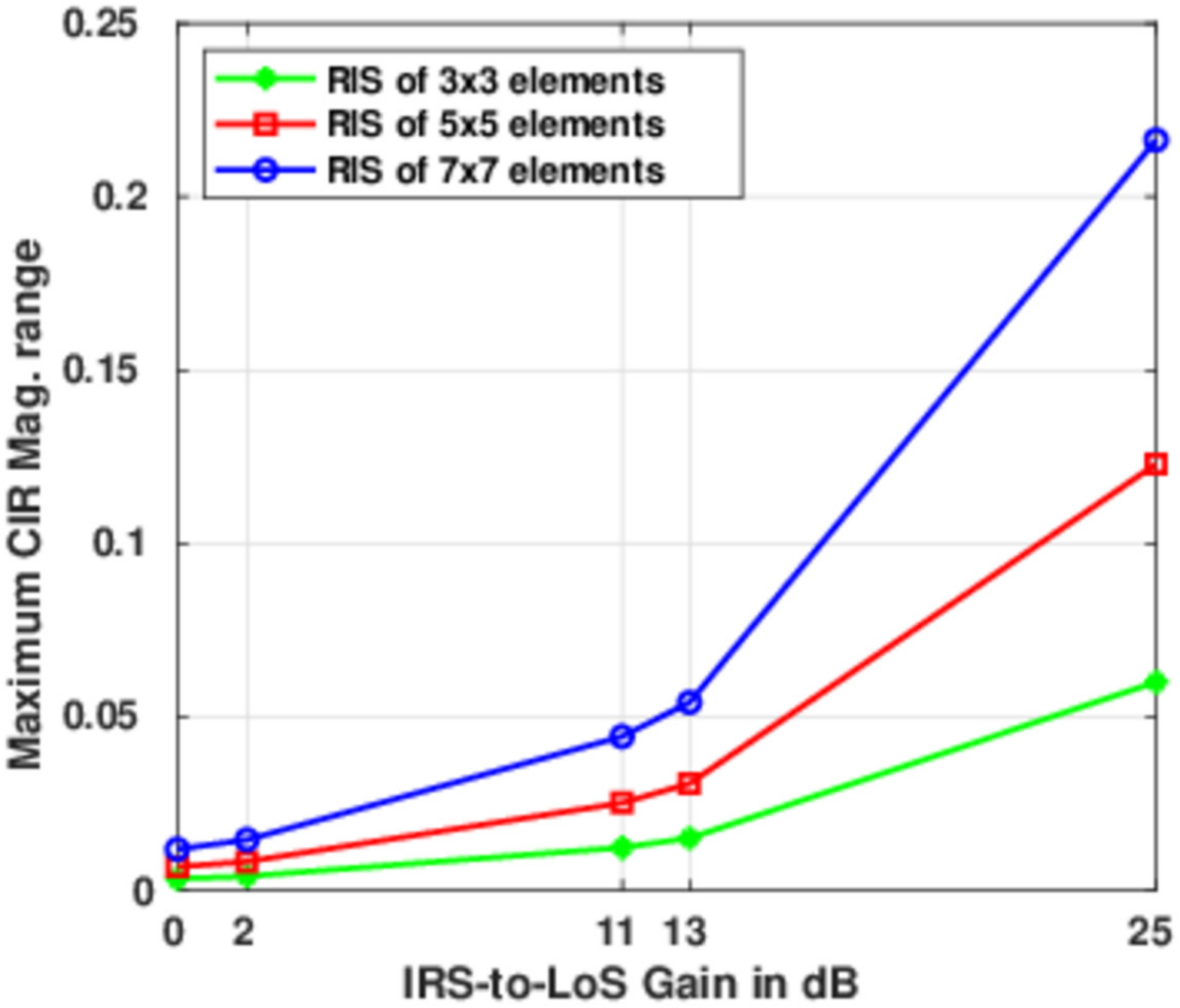}
     \vspace{-16.75mm}
  \caption{CIR range}
  \label{fig:Cir_range}
\end{subfigure}
\caption{\small (a) Higher antenna directivities tend to lead to higher RIS-to-LoS power ratios (difference between blue and red arrows), which translates into (b) higher achievable CIR magnitude range, as the LoS component does not neglect the RIS contribution.}
\label{fig:directTOT}
\vspace{-2mm}
\end{figure}

\vspace{-2mm}
\subsection{Multi-Antenna Relay (Ch4)}  \label{subsec:multi_antenna}
The straightforward implementation of FIR filter taps in the CNN requires (i) constant inter-path time arrivals from consecutive RIS paths, i.e., $t_{RIS_{i+1}}- t_{RIS_i} = \Delta_t, \forall{i} \in \{0,...N-1\}$, and (ii) exact match between these inter-path time arrivals and the communication symbol time, i.e., $\Delta_t=Ts$. 
Here, the first condition imposes a hard constraint on the physical deployment of RIS in the environment, forcing all RIS paths
lengths to be \textit{exact} multiples of one another. 
To achieve this high (sample-level) precision, AirNN accommodates a software-based temporal adjustment over the transmitted frames, as we discuss in Sec.\ref{subsec:airnnnos}.
The second condition requires sampling rates ($Fs=1/Ts$) that may not be compliant with the expected rate at the relay receiver \texttt{R-Rx}.
For example, for a total separation of 2m between two signal paths, the arrival time difference is 66.7~ns, which needs a sampling rate of up to 150 MS/s. AirNN solves this via a multi-antenna system at the \texttt{R-Tx}, where each relay antenna transmits with a time delay of precisely one sample with respect to the next, which maintains equal spacing between arriving signals. For instance, with $Fs = 1$ MS/s, we create a convolution output sample per microsecond if all signal paths are equal in traversed distance. 

\vspace{-2mm}
\subsection{Directional Antennas (Ch5)} \label{subsec:direct_antenna}
\begin{figure}[t!]
\centering
\begin{subfigure}{0.49\linewidth}
  \centering
  \includegraphics[trim=0 0 20 5,clip,width=\linewidth]{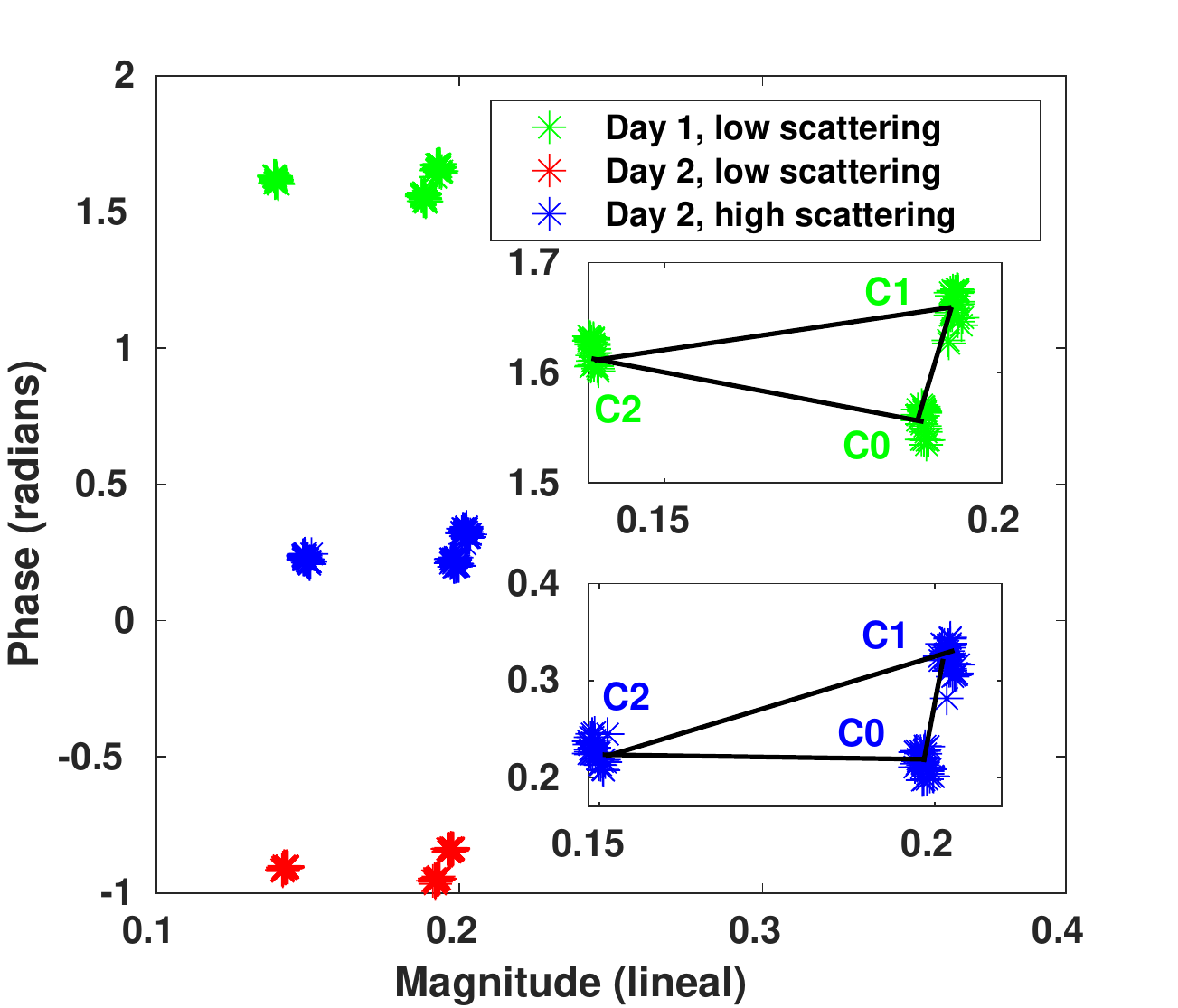}
  \caption{}
  \label{fig:HUncorrected}
\end{subfigure}
\begin{subfigure}{0.49\linewidth}
  \centering
\includegraphics[trim=0 0 20 5,clip,width=\linewidth]{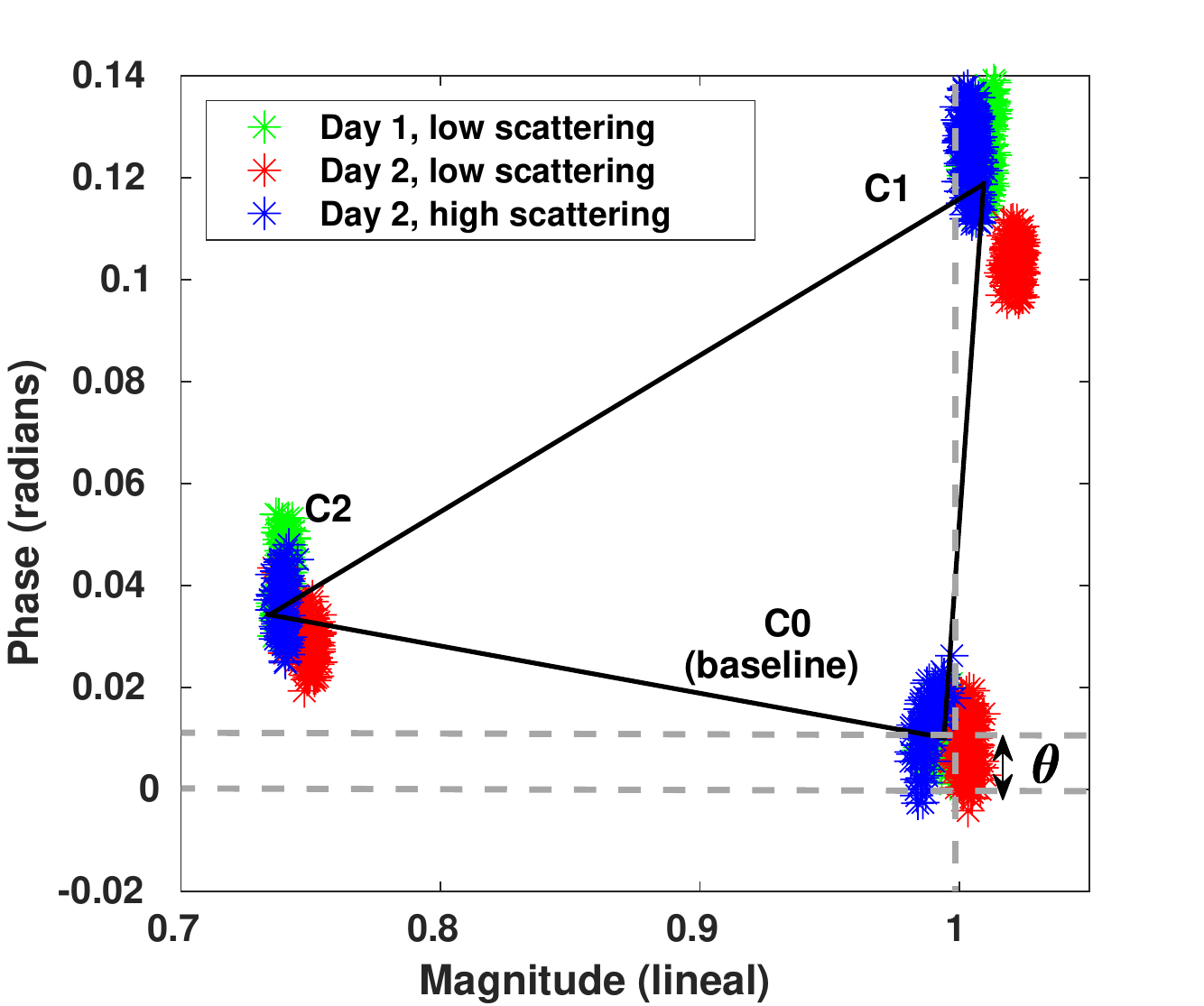}
  \caption{}
  \label{fig:HCorrected}
\end{subfigure}
\vspace{-1mm}
\caption{\small (a) Measured CIR using a single RIS with three phase configurations (C0, C1, and C2) under varying scattering profiles. We observe a relatively similar relative distance between the magnitude and phase plot of the CIR. (b) AirNN adapts to varying conditions using LS  equalization with respect to \textit{a prori} chosen baseline C0.}
\label{fig:figureOffset}
\end{figure}

While the multi-antenna \texttt{R-Tx} ensures fine-grained temporal separation of the signal paths, an omnidirectional transmission will also be reflected from multiple \textit{uncontrolled} scatterers (other than our RIS), present in the environment.
For an omnidirectional transmission, the received signal from a RIS  roughly drops at least 10 dB that in turn reduces the amplitudes of the FIR filter taps. 
In order to study this problem, we formally express the delay profile in a setup with a single antenna \texttt{R-Tx}, single antenna \texttt{R-Rx} and $N$ RIS as:
\begin{equation}
S(t)= (P_t-L_{LoS} )\delta(t_{LoS})+\sum_{i=1}^N (P_t - L_{RIS_i})\delta(t_{RIS_i}),
\label{eq:delayProf}
\vspace{-1mm}
\end{equation}
with $P_t$ (dBm) as the transmitted power. $L_{LoS}$ and $L_{RIS_i}$ (dB) are the losses for the LoS path and the $i^{th}$ RIS path, $i=\{1,2,..., N\}$, respectively. 
Following the interpretation of a given RIS as an array of diffuse reflective antennas \cite{ozdogan2019intelligent}, and considering that each RIS is formed by $M$ such antennas, we estimate $L_{RIS_i}$  from:
\vspace{-2mm}
\begin{equation}
L_{RIS_i} = 10log \mid\sum_{m=1}^M l_{RIS_i^m} e^{j\phi_i^m}
\mid,
\label{eq:scattererToIRS}
\vspace{-1mm}
\end{equation}
where $l_{RIS_i^m}$ represents the path loss associated with a particular reflective antenna $m$. This loss value depends on the carrier frequency, the distance between \texttt{R-Tx} to RIS and RIS to \texttt{R-Rx}, RIS dimensions, \texttt{R-Tx} and \texttt{R-Rx} antenna gain in the direction of each reflective antenna, and the angle of incidence of the signal wavefront to the RIS plane. The term $e^{j\phi_i^m}$ in Eq.\ref{eq:scattererToIRS} represents the phase of the incoming signal from RIS element $m$ to the \texttt{R-Rx}, and thus, the received power is determined by the interference of the incoming signal from all $m=\{1,2,..., M\}$ reflective antennas of the RIS. The phase $\phi_i^m$ is given by $\phi_i^m = k(d_{(R-Tx,_i^m)}+d_{(_i^m,R-Rx)}
)+\phi_{S_i^m}$
with $k=\frac{2\pi}{\lambda}$ as the wave number and $\lambda$ as the wavelength. Lastly, $d_{(R-Tx,_i^m)}$, $d_{(_i^m,R-Rx)}$ represent the distance from \texttt{R-Tx} to reflective antenna $m$ and from that same antenna to \texttt{R-Rx}, respectively. The term $\phi_{S_i^m}$ gives the configurable phase shift introduced by the reflective antenna $m$. Importantly, the estimation of $l_{RIS_i^m}$ follows a \textit{product-distance} path loss model \cite{ozdogan2019intelligent}, where the power decays with $d_{(R-Tx,_i^m)}d_{(_i^m,R-Rx)} \sim d^2$ in contrast to $d$, in conventional beamforming. 
We estimate the term $L_{LoS}$ in Eq.\ref{eq:delayProf} from the Friis equation and the delay terms $\delta(t_{LoS})$ and $\delta(t_{RIS_i})$ by dividing the known distances with $c$, the speed of light in vacuum.
Thus, we estimate the CIR component associated to the RIS $i$ path as:
\vspace{-2mm}
\begin{equation}
h_i = \sqrt{l_{RIS_i}}\sum_{m=1}^M  e^{j\phi_i^m}.
\label{eq:CIRtheory}
\end{equation}

In Fig.~\ref{fig:Cir_range}, we show the simulated maximum range for the magnitude of the received signal given in Eq.\ref{eq:CIRtheory} as a function of the RIS-to-LoS power ratio, for different \texttt{R-Tx} antenna radiation patterns (Fig.~\ref{fig:directTOT}a), as defined by their respective 3-dB beamwidth $BW_{3dB} = \{360\degree,120\degree,50\degree,25\degree,12\degree\}$ and different number of antenna elements with $M = \{49,36,25,9,4\}$. In Fig.~\ref{fig:directTOT}a, the blue arrow points to the RIS, while the red arrow points directly to the \texttt{R-Rx}, located at $45\degree$ from the RIS direction. We observe that for low antenna directivity, the high power of the LoS component compared to that of the signal reflected from the RIS renders any manipulation of the RIS ineffective. Hence, AirNN uses directional antenna elements at the \texttt{R-Tx} that (i) boost the power of the reflected signals from the RIS paths and (ii) mitigate the degrading impact of the LoS signal along with the effect of additional ambient scatterers not controlled within AirNN.

\subsection{Compensating for Channel (Ch6)}  \label{CIR_offset}

\begin{figure}[t]
    \centering
   \includegraphics[trim=0 280 0 300, clip, width=1.2\linewidth]{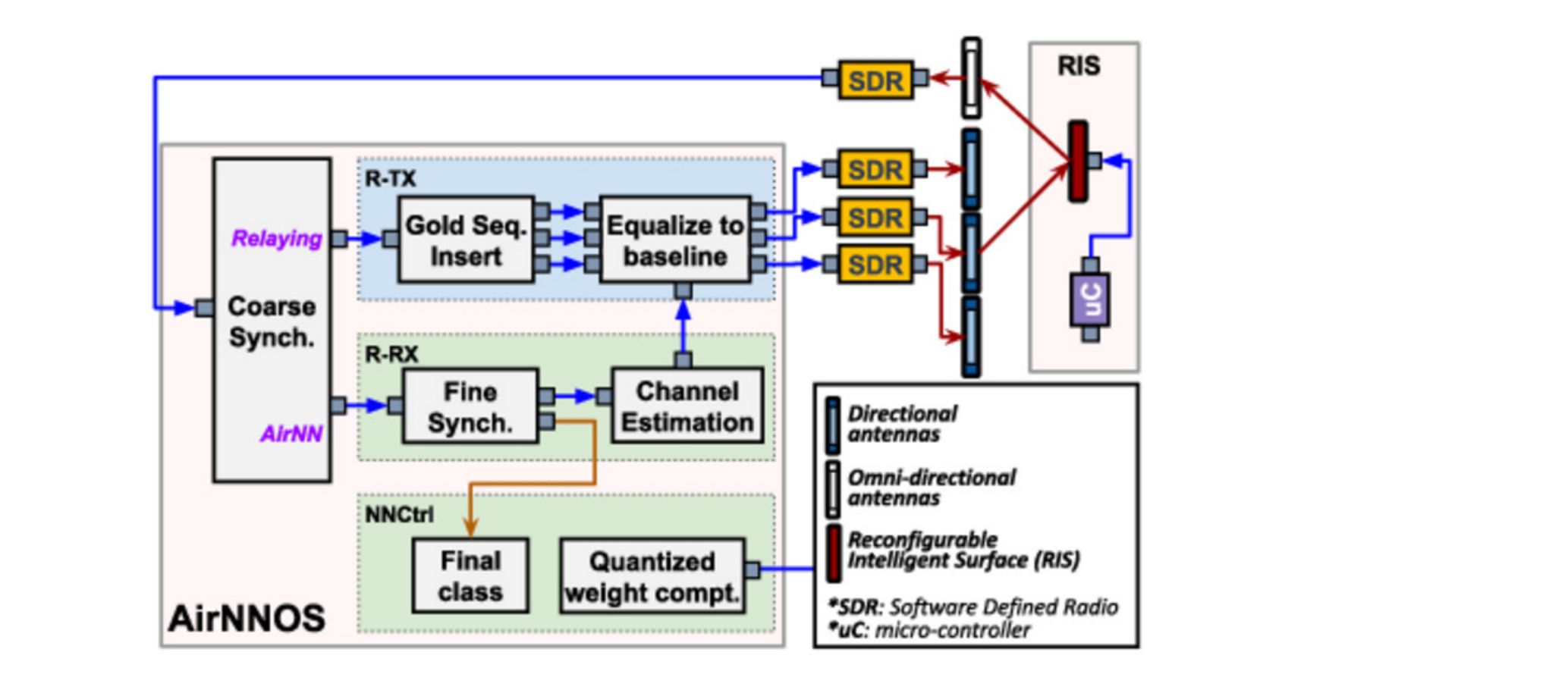}
    \caption{\small AirNNOS software blocks within the relay modules of  \texttt{R-Tx}, \texttt{R-Rx} and \texttt{NNCtrl}.}
    \centering
\label{fig:AirNN_system_blocks}
\end{figure}

Due to the non-stationarity of wireless channels, the CIR engineered by certain RIS configurations may change over time, not resulting in the exact weights corresponding to the digital convolution, unless the CNN architecture is re-trained for every new scattering profile.
Instead, AirNN uses a channel tracking and correction stage to ensure that the weights of the CNN, as decided by the RIS configuration, remain valid even under new channel conditions. This ensures that the received signal at the \texttt{R-Rx} always experiences a fixed and constant phase of zero degrees and unit magnitude when using the baseline configuration at every RIS.  

We explain this process in Fig.~\ref{fig:figureOffset} as follows: we first collect signal samples using the standard AirNN relay setup. 
We then compute the phase rotation for three RIS configurations, i.e., $C0$, $C1$, and $C2$, on two different days and scattering profiles.
These profiles include cases of low impact (few meters away) and high impact (few cm away) scatterers, respectively. 
We use Least Squares (LS) channel estimation with a known preamble sequence. 
Since directional transmissions at \texttt{R-Tx} mitigate the multipath effect, we denote $\hat{\textbf{h}}_{C0} \in \mathbb{C}$ as the estimated narrowband channel at \texttt{R-Rx}.
First, we observe that a change in the channel environment introduces variations within the over-the-air generated filter coefficients (Fig.~\ref{fig:HUncorrected}). 
Second, the relative distance between any pair of clusters of channel coefficients resulting from specific RIS configurations remains constant between deployments (Fig.~\ref{fig:HUncorrected}).
In this example, AirNN takes configuration $C0$ as the baseline to generate the desired unit magnitude and no phase rotation, as shown in Fig.~\ref{fig:HCorrected}. To accomplish this, the \texttt{R-Tx} inverts the estimated channel vector as  $\textbf{p}=(\hat{\textbf{h}}_{C0})^{-1} \in \mathbb{C}$ to equalize the channel transformation for a given RIS configuration. We capture such translation by $\theta$ ($0 \leq \theta \leq \pi/2$) computed as $\cos(\theta) = \textit{Re} \left\lbrace \textbf{h}^{H} \textbf{p} \right\rbrace / (||\textbf{h}|| ||\textbf{p}||)$. 

\begin{figure}[t]
    \centering
    \includegraphics[trim=30 50 280 40,clip,width=0.8\linewidth]{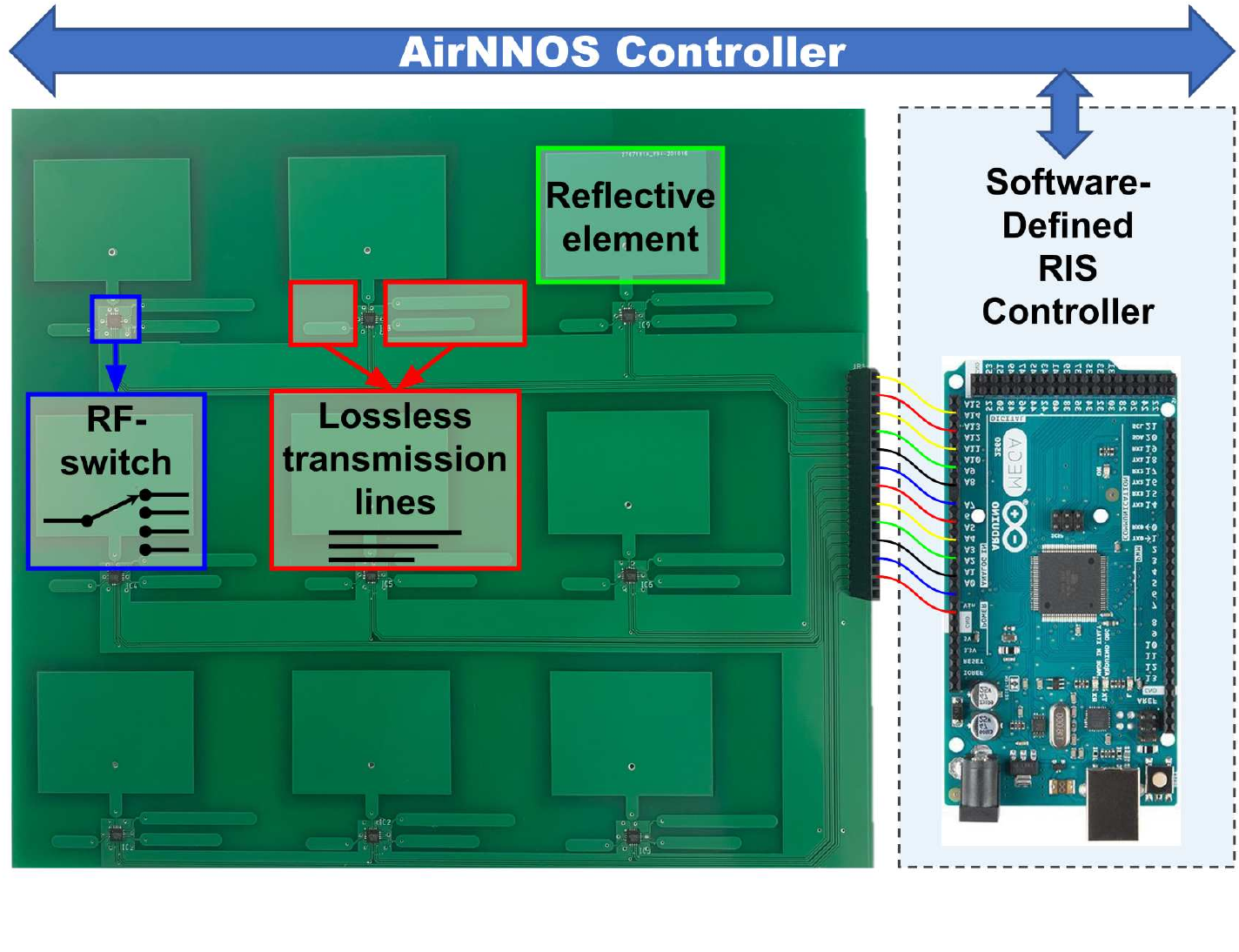}
    \caption{Hardware prototype of a RIS with 9 patch reflective antennas whose signal reflections can be changed through software running in the controller by selecting transmission lines via the RF-switch.}
    \centering
\label{fig:IRSDesign1}
\end{figure}

%% file: sections/setup.tex
\section{System Implementation}  \label{systemArch}
We highlight the main components of AirNN from a hardware viewpoint in  Fig.~\ref{fig:AirNN_system_blocks}. We describe the implementation of the relay node using COTS Software Defined Radios (SDRs) in Sec.~\ref{subsec:airnn_hw_components} and the RIS hardware design and implementation in Sec.~\ref{subsec:RIS_implementation}. We describe AirNNOS software that drives the relay operations and controls the RIS units in  Sec.~\ref{subsec:airnnnos}.  

\subsection{Relay Hardware Components}  \label{subsec:airnn_hw_components}
\noindent $\bullet$ \textbf{SDRs:} In our implementation, the front-end of the relay is composed of four Ettus USRP X310 SDRs, each attached to a UBX 160MHz daughterboard, which can flexibly digitize up to 200 MSamples per second. Three SDRs serve in the \texttt{R-Tx} interface, and a single SDR serves in the \texttt{R-Rx} interface. The SDRs are connected to the host machine via a 1 Gbps Ethernet link. We synchronize all SDRs in frequency and time through an Ettus OctoClock CDA-299. The octoclock helps correct the CFO of the multiple \texttt{R-Tx} and the \texttt{R-Rx}.

\noindent $\bullet$ \textbf{Antennas:} The relay receiver \texttt{R-Rx} is attached to a VERT 2450 dual-band omnidirectional vertical antenna with 3 dBi gain. The relay transmitters \texttt{R-Tx} have directional patch antennas with 18$\degree$ of 3-dB beamwidth in azimuth and elevation and a grating lobe at 90$\degree$ from their broadside direction. They operate in the 2.4GHz band.

\input{sections/RIS_tentative}

\subsection{AirNNOS Controller (Ch7)}  \label{subsec:airnnnos}

Our orchestrating software framework called AirNNOS controls the following (see Fig.~\ref{fig:AirNN_system_blocks}):

\noindent$\bullet$ \textbf{Transmission/reception sequence:} 
\texttt{R-Rx}  collects IQ samples and forwards them to the \textit{Coarse synch} block that performs basic energy detection. This triggers AirNN relay actions (see \textit{Relaying} output in Fig.~\ref{fig:AirNN_system_blocks}). 
At this stage, the \textit{Coarse synch} module redirects the incoming IQ samples to the associated \texttt{R-Tx} thread,  which in turn processes the samples for AirNN operation and then re-transmits over the air. At this time, the \textit{Coarse synch} switches its active output port to forward the incoming samples, i.e., resulting from the over-the-air convolution, to the processing blocks within the \texttt{R-Rx} (see \textit{AirNN} output). 
After fine-grained synchronization at the \texttt{R-Rx}, samples are fed to the \texttt{NNCtrl} to complete the CNN processing.

\noindent$\bullet$\textbf{Pre-processing at the \texttt{R-Tx}:}
\begin{figure}[t!]
\centering
\begin{subfigure}{0.49\linewidth}
  \centering
  \includegraphics[trim=0 0 20 0,clip,width=\linewidth]{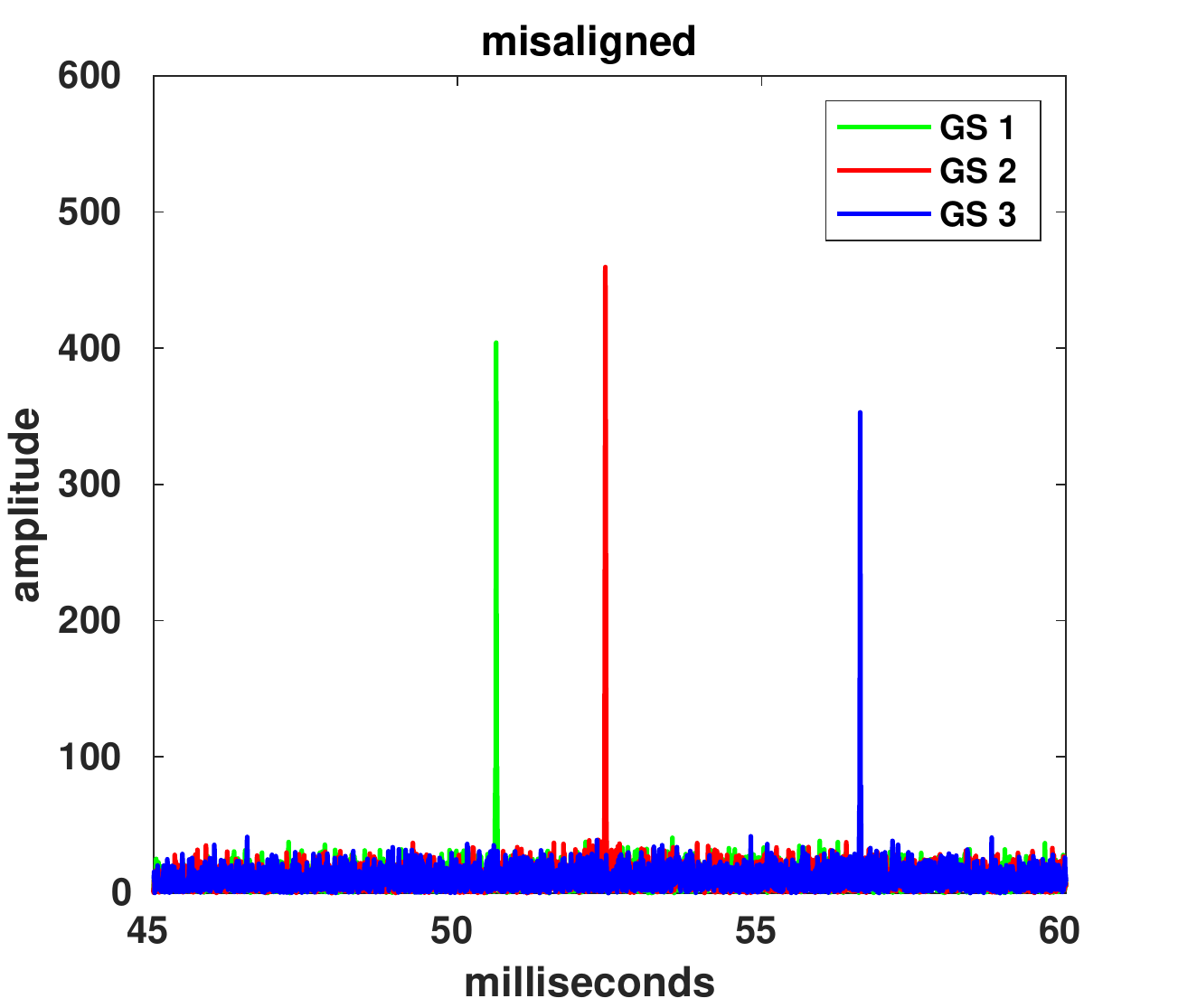}
  \caption{}
  \label{fig:Notalignment}
\end{subfigure}
\begin{subfigure}{0.49\linewidth}
  \centering
\includegraphics[trim=0 0 20 0,clip,width=\linewidth]{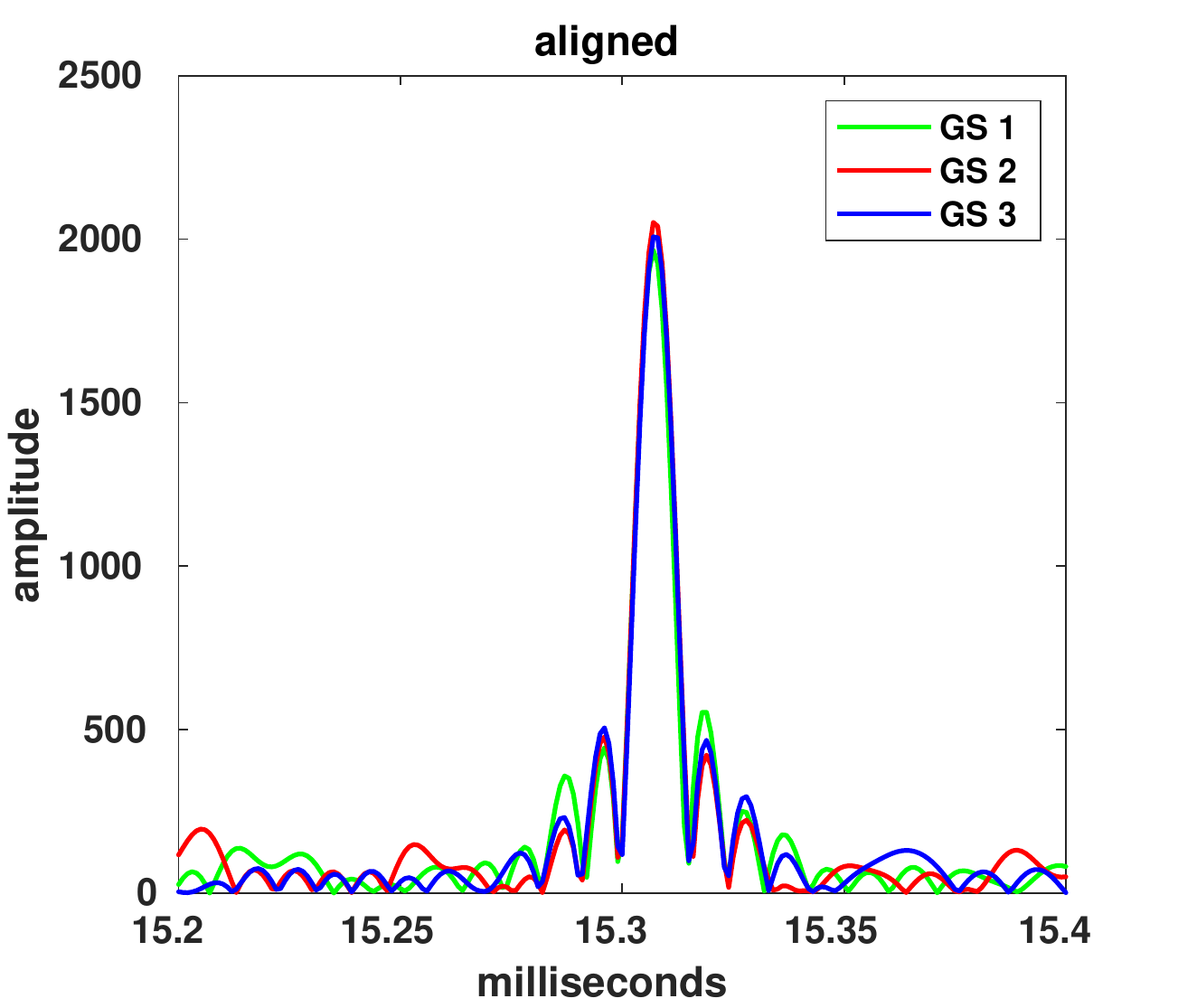}
  \caption{}
  \label{fig:alignment}
\end{subfigure}
\caption{\small Precise synchronization at the \textit{R-Tx}. (a) Misaligned inter-RIS path arrivals, and (b) precise synchronization in AirNN (see Sec.~\ref{subsec:challenges}).}
\end{figure}
The \texttt{R-Tx} generates a set of orthogonal Gold sequences (GS) \cite{Xinyu2011AnalysisSystem} that have desirable properties of good auto- and cross-correlation, and uniquely assigns one sequence to the set of IQ samples being sent over each transmit antenna. Thus, a transmission from the relay is composed of a GS appended to the received samples from the Tx. 
The benefit of using GS is two-fold: 
on the one hand, GS guarantees precise time synchronization for  generating paths-delays that match the temporal distribution of desired FIR filters (see Sec.~\ref{subsec:multi_antenna}).
On the other hand, GS offers a way to estimate and compensate for the channel variations over time (Sec.~\ref{CIR_offset}), as we explain next.

\noindent $\bullet$ \textbf{Synchronization and channel estimation at the  \texttt{R-Rx}:}
Although the SDRs in the relay \texttt{R-Tx} start their transmissions at the same PPS instant, AirNN requires more fine-grained precision to realize the desired filter taps (Fig.~\ref{fig:Notalignment}) when transmitting with high bitrate.
To achieve this, the \texttt{R-Rx} computes the symbol misalignment between all the transmit streams via the GS by setting the last received stream as a reference. It then sends back this information along with the channel state information or CSI (\textit{Channel estimation} computed in the \texttt{R-Rx}) to the \texttt{R-Tx}. 
The SDRs in the \texttt{R-Tx} delay their signals with additional zero-padding to sync with other peer-transmitters.
Only with accurate time alignment (Fig.~\ref{fig:alignment}) can AirNN generate the desired temporal displacement by deferring transmissions precisely by one sample with respect to other transmitters (see Sec.\ref{subsec:multi_antenna}).

%% file: sections/RIS_tentative.tex
\vspace{-2mm}
\subsection{RIS Hardware Design and Fabrication} \label{subsec:RIS_implementation}

The concept of loss-based transmission line for phase shifting has been implemented before \cite{dunna2020scattermimo}, which we modify to realize AirNN. Our fabricated RIS is shown in Fig. ~\ref{fig:IRSDesign1}. 

\textcolor{black}{To access the feasibility of generating over-the-air FIR weights, we next study several RIS parameters, including (i) the type and number of patch reflective antennas within a RIS, the intra-RIS separation of the reflective antennas, and (ii) the phase shifts that these antennas may generate.
We leverage the signal propagation model presented in Sec.~\ref{subsec:direct_antenna} to assess the impact of several RIS  parameters on the feasible over-the-air generated FIR weights. We include simulations using a topology of a single RIS placed equidistantly from the \texttt{R-Tx} and \texttt{R-Rx} antennas at 2.5 meters, while \texttt{R-Tx} and \texttt{R-Rx} antennas are separated by 5 meters.}

\begin{figure}[t!]
\centering
\begin{subfigure}{0.49\linewidth}
  \centering
  \includegraphics[trim=0 0 0 0,clip,width=\linewidth]{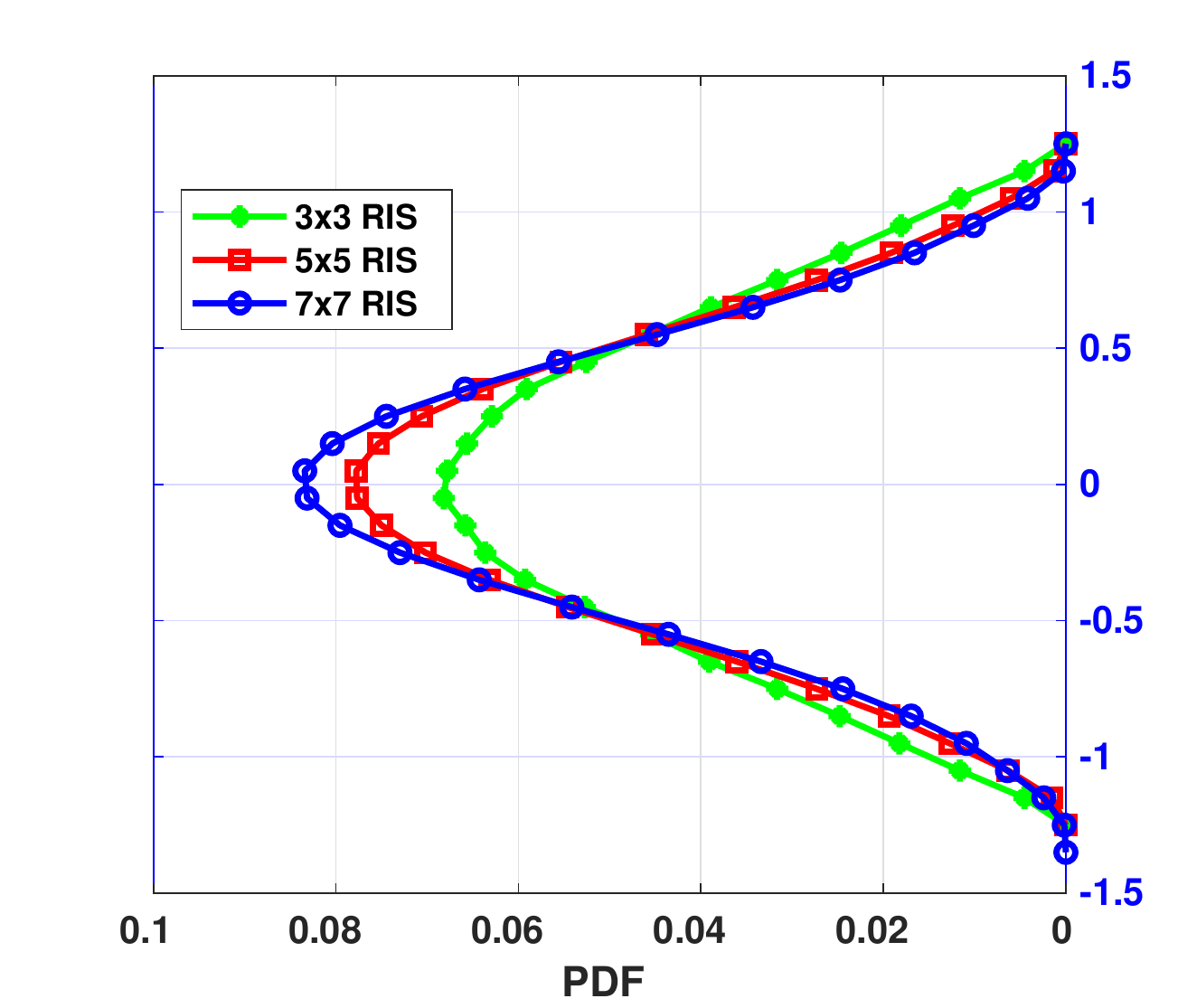}
  \caption{}
  \label{fig:phaseNantenna}
\end{subfigure}
\begin{subfigure}{0.49\linewidth}
  \centering
\includegraphics[trim=0 0 0 0,clip,width=\linewidth]{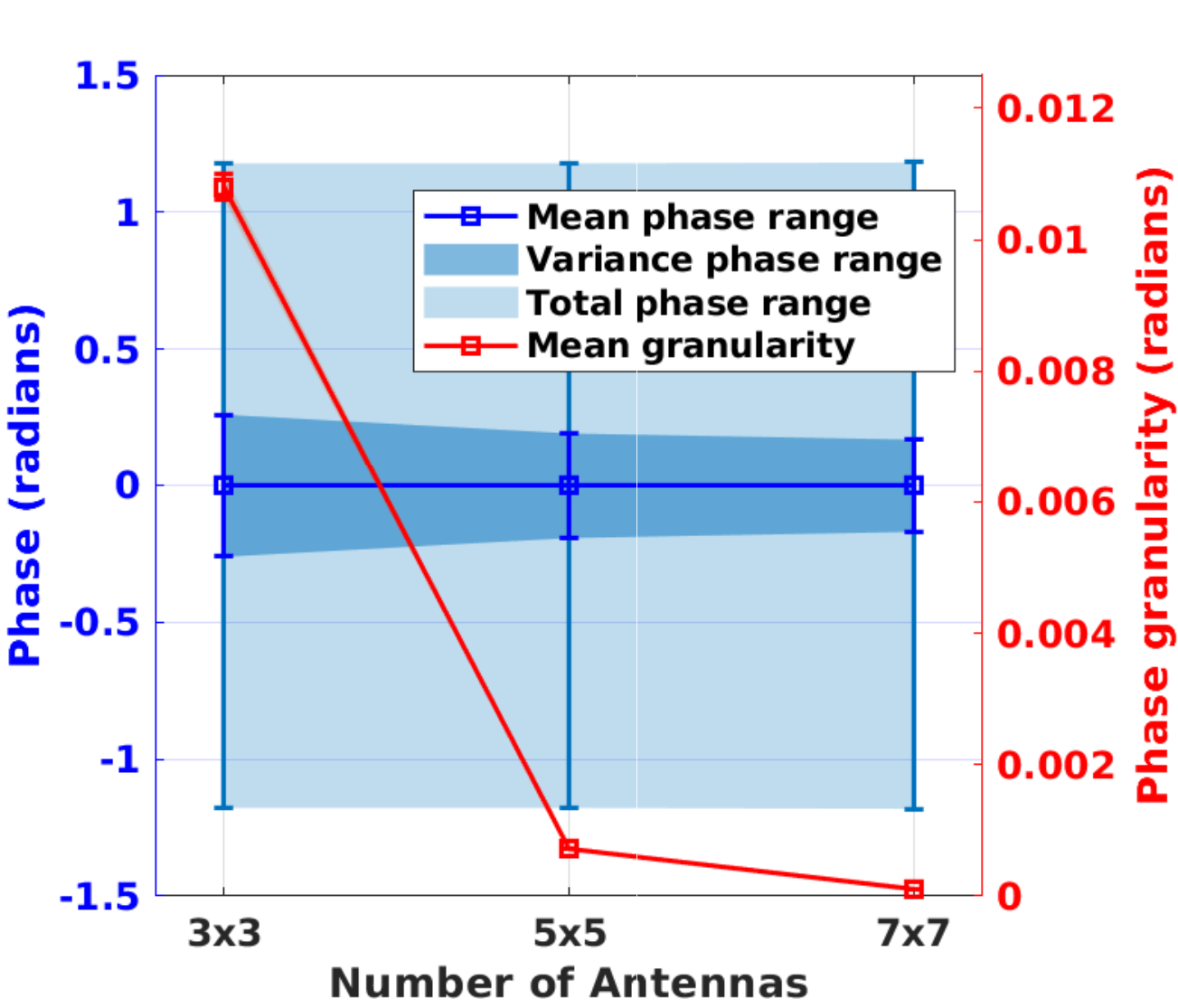}
  \caption{}
  \label{fig:phaseGran}
\end{subfigure}
\caption{\textcolor{black}{CIR phase range and distribution w.r.t. RIS dimensions (i.e., number of reflective antennas). (a) A larger dimension does not result in a larger CIR phase range, but (b) results in a better granularity (less CIR weight quantization).}}
\label{fig:analysis_RIS_dimensions}
\end{figure}

\noindent $\bullet$\ \textbf{Reflective Elements: (type, size, distribution and number):%
}
Each reflective element within our RIS is a switchable patch-type antenna of dimension $\lambda$/2 inserted between a two-layer PCB dielectric substrate and a full metal sheet at the bottom layer. We select a \textcolor{black}{RIS} of $M=9$ reflective antennas, with a 3x3 layout in a 2D-plane.
We space the antennas a distance of $\lambda$/2 to reduce the effect of mutual coupling between neighboring elements, as well as grating lobes in the RIS radiation pattern \cite{balanis2016antenna}.

To assess how AirNN can benefit from having a larger number of reflective antennas, we configure our simulations with three different RIS sizes, i.e., 3x3, 5x5 and 7x7, and evaluate the obtained phase span, as well as the resulting phase granularity.
Here, the term \textit{span} refers to the difference between the maximum and the minimum induced phase shifts possible at the \texttt{R-Rx}, whereas the term \textit{granularity} refers to the minimum phase difference between any two realizable phases at the \texttt{R-Rx}.
\textcolor{black}{Interestingly, having a larger RIS dimension does not lead to a larger span, as we show in Fig.~\ref{fig:phaseNantenna}. 
Although a larger number of reflective elements achieve higher granularity, as seen in Fig.~\ref{fig:phaseGran}, this improvement increases the size and cost of the RIS. For example, the improvement in granularity obtained by using a 7x7 instead of a 3x3 antenna RIS increases the manufacturing price by $\$200$.
These findings motivate us to select a small antenna set of $M=9$ reflective antennas, with a 3x3 layout in a 2D-plane.}
\textcolor{black}{These antennas are finally printed on a RIS PCB board with a FR-4 epoxy glass substrate of dimension $20$cm $\times$ $20$cm $\times$ $0.16$cm.}

\noindent \textcolor{black}{$\bullet$\ \textbf{Phase Shifts (angular range and inter-shift distances)}:}
 We connect each of the nine reflective antennas to three loss-less transmission lines of different lengths through a single RF switch. The resulting four phases per antenna (including no phase shift) enables $(4)^9$ configurations per RIS, generating a rich diversity of distinct signal reflections at the \texttt{R-Rx}. This design is finally printed on a RIS PCB board with a FR-4 epoxy glass substrate of dimension $20$cm $\times$ $20$cm $\times$ $0.16$cm. A general purpose HMC7992 RF switch \cite{rfswitch} connected to an Arduino Mega2560 $\mu$controller activates the selected line per antenna in real-time. 

\textcolor{black}{By selecting the length of our transmission lines, we can alter each reflective element impedance, which in turn changes their reflective coefficient and, consequently, introduces a phase shift to the signal reflected by that particular element.} \textcolor{black}{Our implementation allows four possible shifts per element, although the overall phase at the receiver is a combination of the individual shifts introduced by each of these nine reflective antennas. }

\begin{figure}[t!]
\centering
\begin{subfigure}{0.49\linewidth}
  \centering
  \includegraphics[trim=0 0 0 0,clip,width=\linewidth]{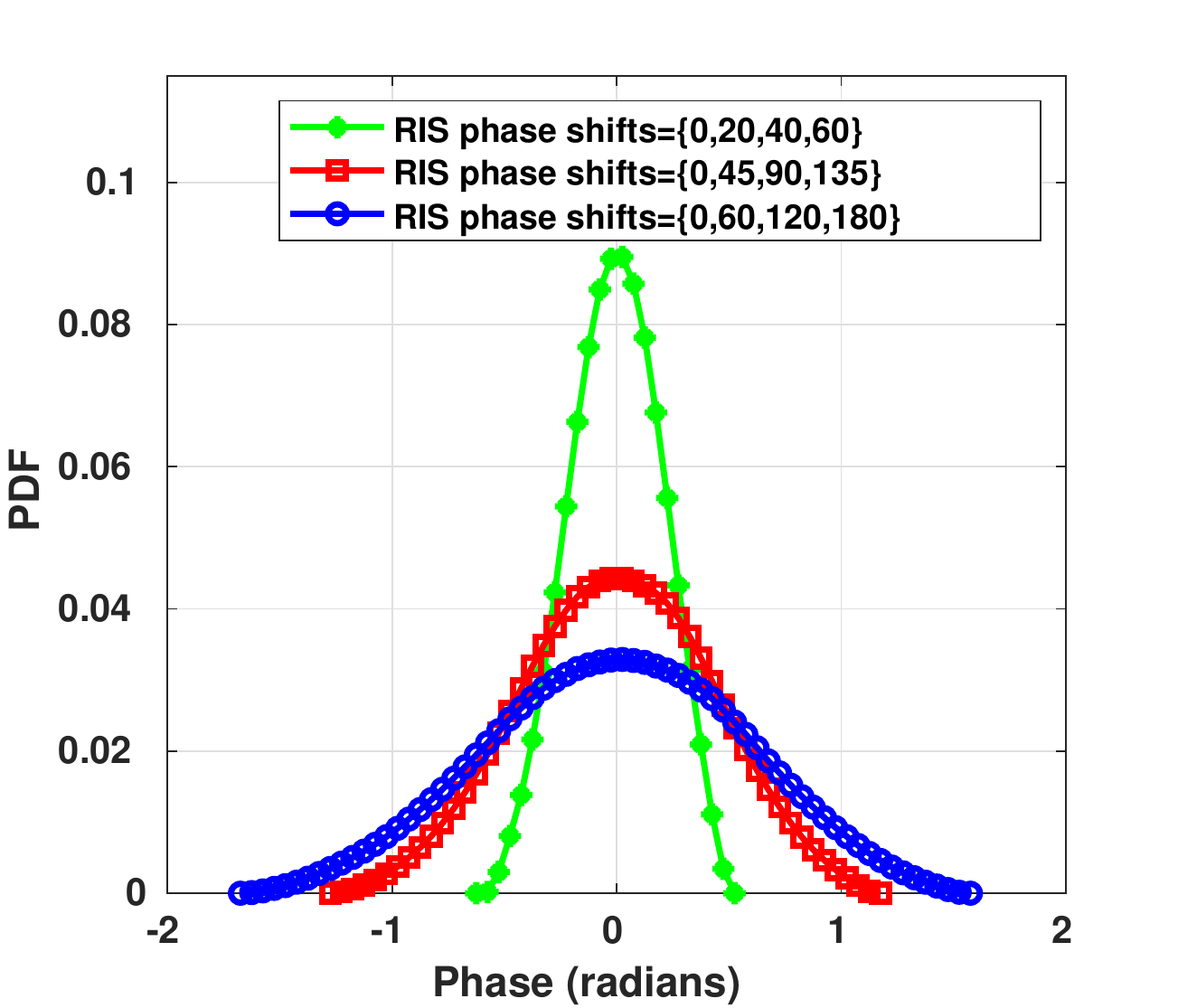}
  \caption{Shift span}
  \label{fig:shiftSpan}
\end{subfigure}
\begin{subfigure}{0.49\linewidth}
  \centering
  \includegraphics[trim=0 0 0 0,clip,width=\linewidth]{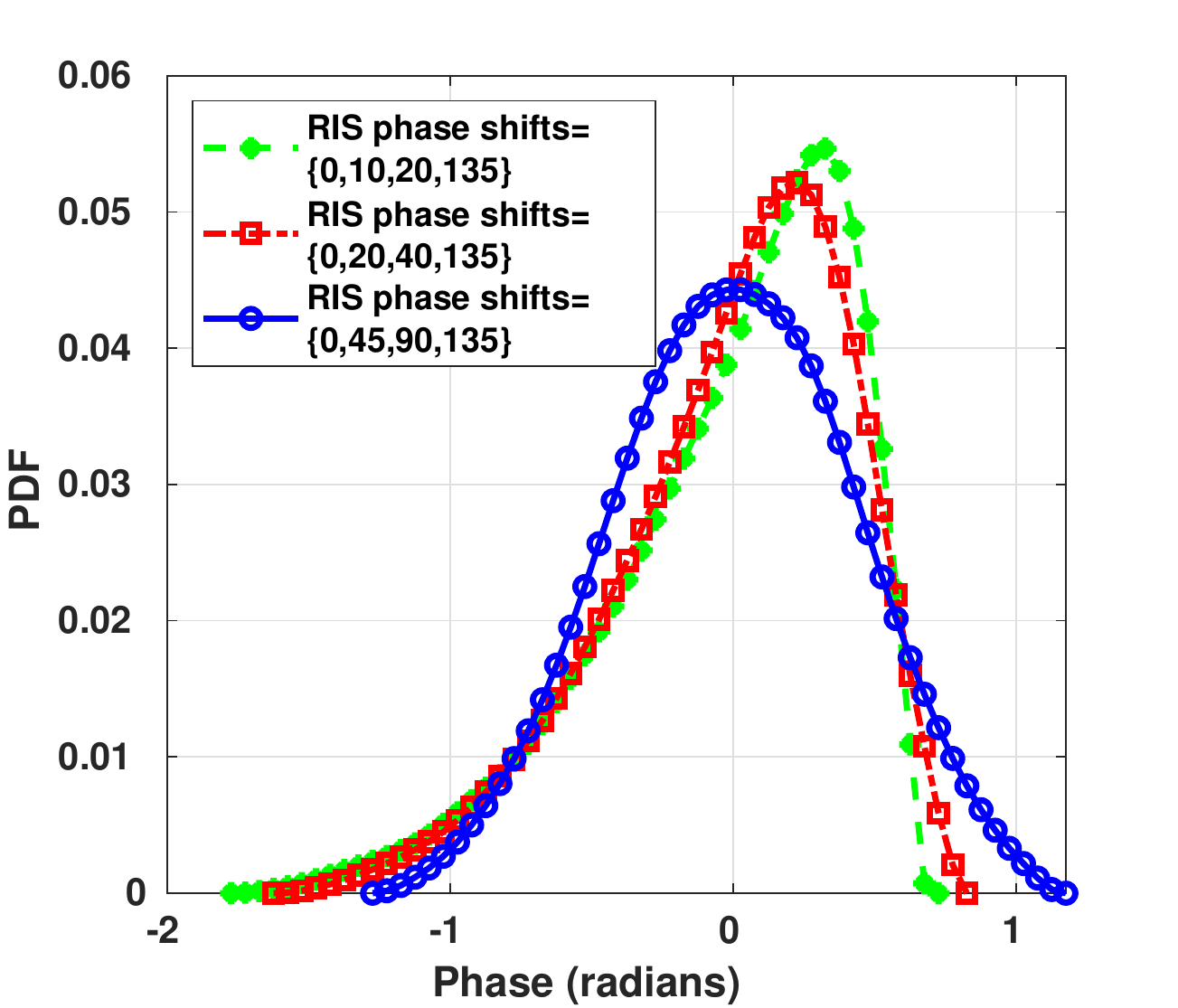}
  \caption{Shift distribution}
  \label{fig:shiftDist}
\end{subfigure}
\caption{\textcolor{black}{Study of the CIR phase range and distribution w.r.t. the RIS. (a) A larger shift span achieves a broader range of CIR phase values. (b) A uniform inter-shift spacing results in a uniform CIR phase distribution.}}
\end{figure}

\textcolor{black}{
We determine next the \textit{span} of possible phases, considering all possible combinations of the transmission line selections per reflective antenna. To do so, we simulate a single RIS unit of 3x3 dimensions with four transmission lines but with different upper bounds of the maximum transmission line induced shift. We show this analysis in Fig.~\ref{fig:shiftSpan} for such maximum shifts of $60\degree$, $135\degree$, and $180\degree$.
We observe that the span at the \texttt{R-Rx} for the transmission line shift of $180\degree$ is over double that of a lower maximum value of $60\degree$. Next, we study how the inter-shift angular distance shapes the range of realizable phases at the \texttt{R-Rx}. We consider three combinations- with uniformly distributed phase shifts between transmission lines, narrowly spaced, and widely spaced shifts. From Fig.~\ref{fig:shiftDist} we see that the PDF for uniform spacing follows a Gaussian phase distribution. Conversely, non-uniform phase spacing results in a Rician distribution. Thus, we design transmission lines to generate uniformly separated phase shifts to enable a maximum span at the receiver, i.e., $\{45\degree, 90\degree, 135\degree\}$.}

%% file: sections/performanceEval.tex
\section{Performance Evaluation}
\label{perfEval}
This section validates the equivalence between the over-the-air convolution using our proposed AirNN system and its digital counterpart.
Then, using a classical example problem of digital modulation classification, we compare in simulation the classification accuracy achieved via different deep learning approaches. This comparison includes classical CNN trained using standard methods on a GPU (Classical-CNN), quantized CNN (QM-CNN) from Sec.~\ref{subsec:constrainedCNN}, and robust-quantized CNN (RQM-CNN from Sec.~\ref{subsec:dataAug}).

\subsection{Experimental Testbed}
\label{expTestbed}
\begin{figure}[t]
    \centering
    \includegraphics[trim=140 50 180 40, clip, width=0.95\linewidth]{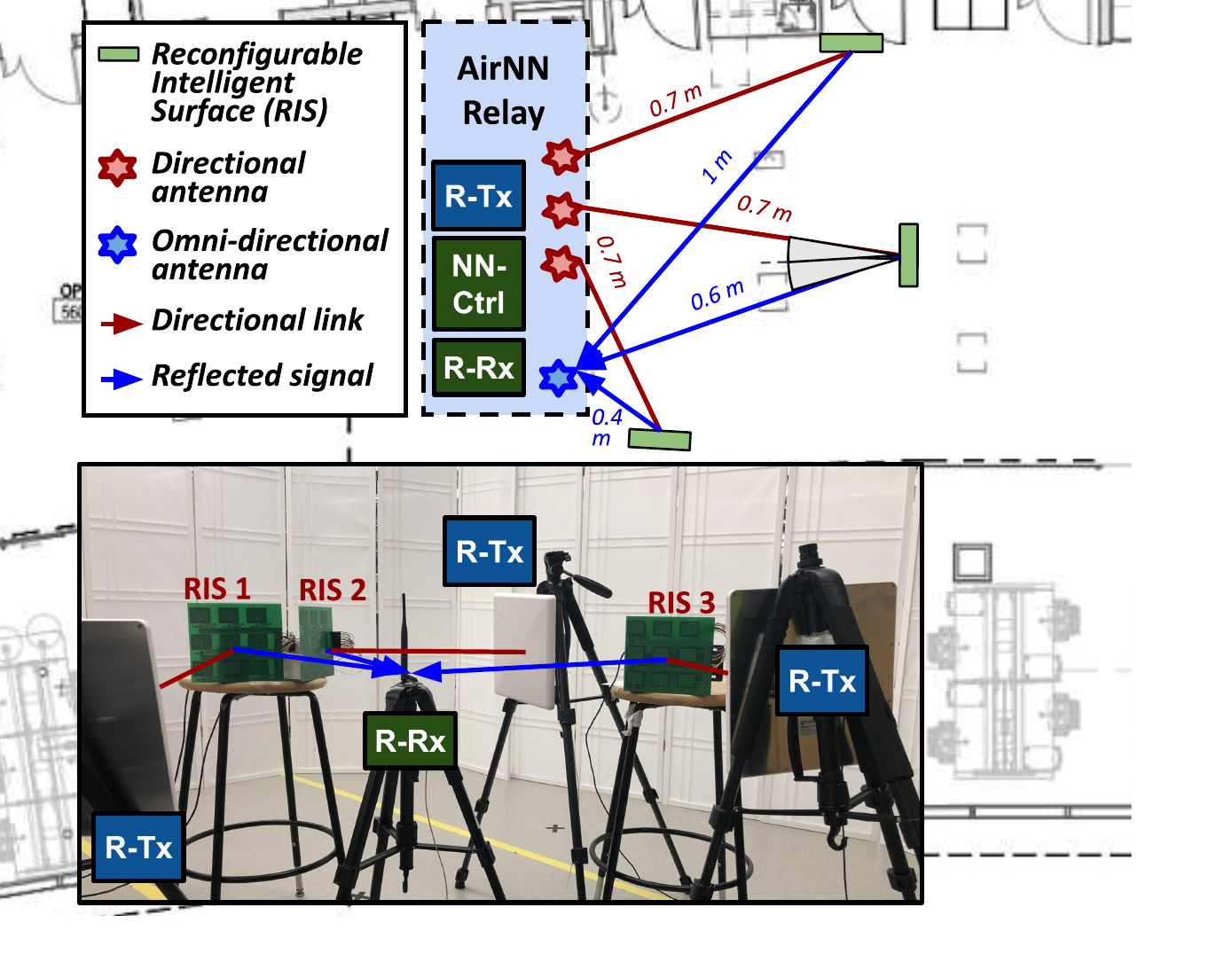}
    \caption{\small Floor layout plan and experimental setup for AirNN testbed using our custom-built RIS. }
    \centering
\label{fig:floorlayout}
\end{figure}

We use the relay as described in Sec.\ref{subsec:airnn_hw_components} and fabricate three RIS units using the approach in Sec.\ref{subsec:RIS_implementation}. The distance between the RIS and both \texttt{R-Tx} and \texttt{R-Rx} are 0.7m. To achieve interference nulls at unintended RIS, given the \texttt{R-Tx} antenna grating lobes, we orient each \texttt{R-Tx} antenna towards a dedicated RIS with a $45\degree$ angle with respect to the plane of any other neighboring RIS (see Fig.~\ref{fig:floorlayout}). We use the frequency of 2.49 GHz and a sampling rate of 1 MS/s. 

\subsection{Validation of Over-the-air Convolution}
\begin{figure}[t!]
\begin{subfigure}{0.485\linewidth}
  \centering
  \includegraphics[trim=30 0 0 200,clip,width=\linewidth]{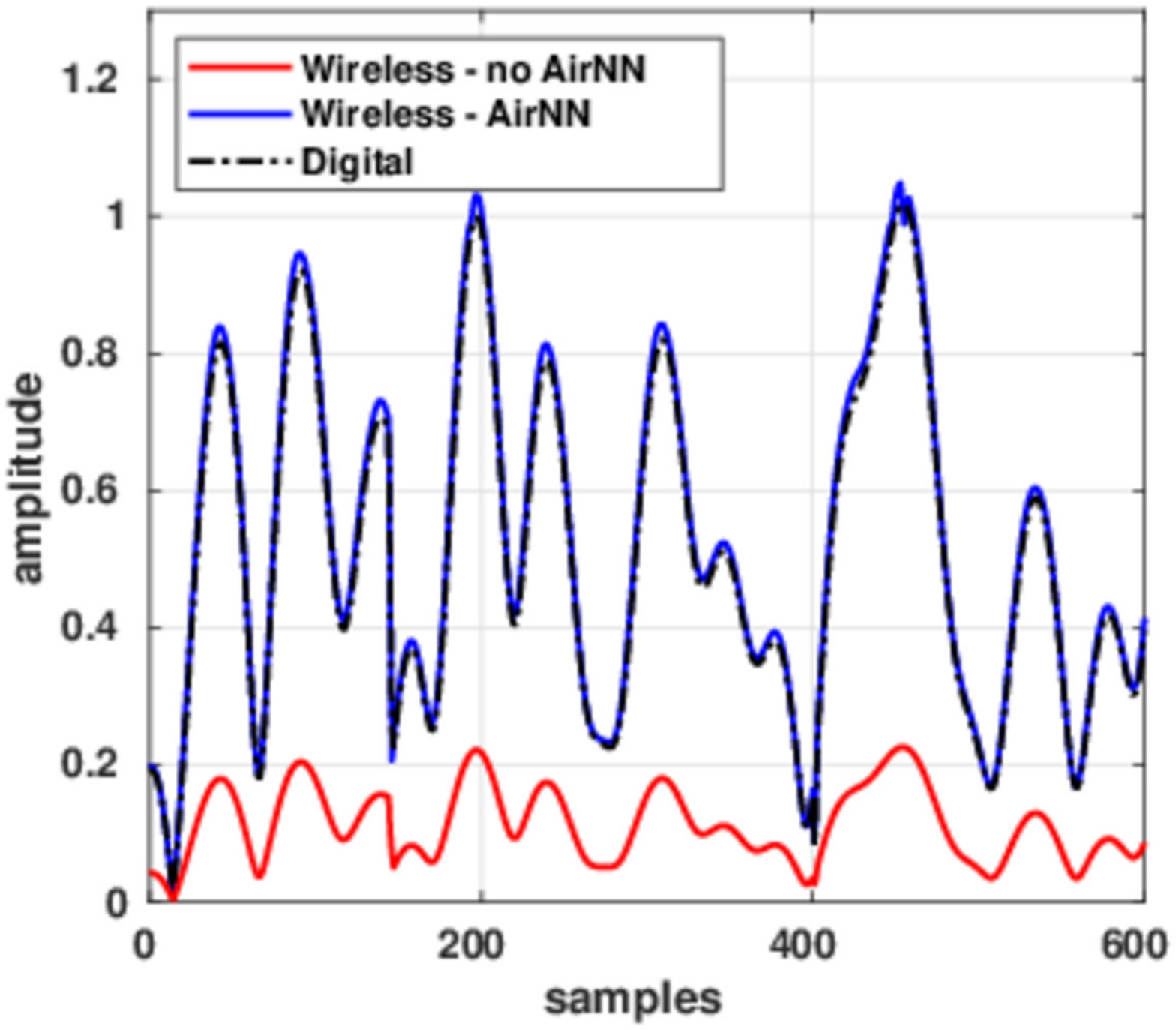}
  \vspace{-17mm}
  \caption{Magnitude}
  \label{fig:magConv}
\end{subfigure}
\begin{subfigure}{0.485\linewidth}
  \centering
  \includegraphics[trim=30 0 0 200,clip,width=\linewidth]{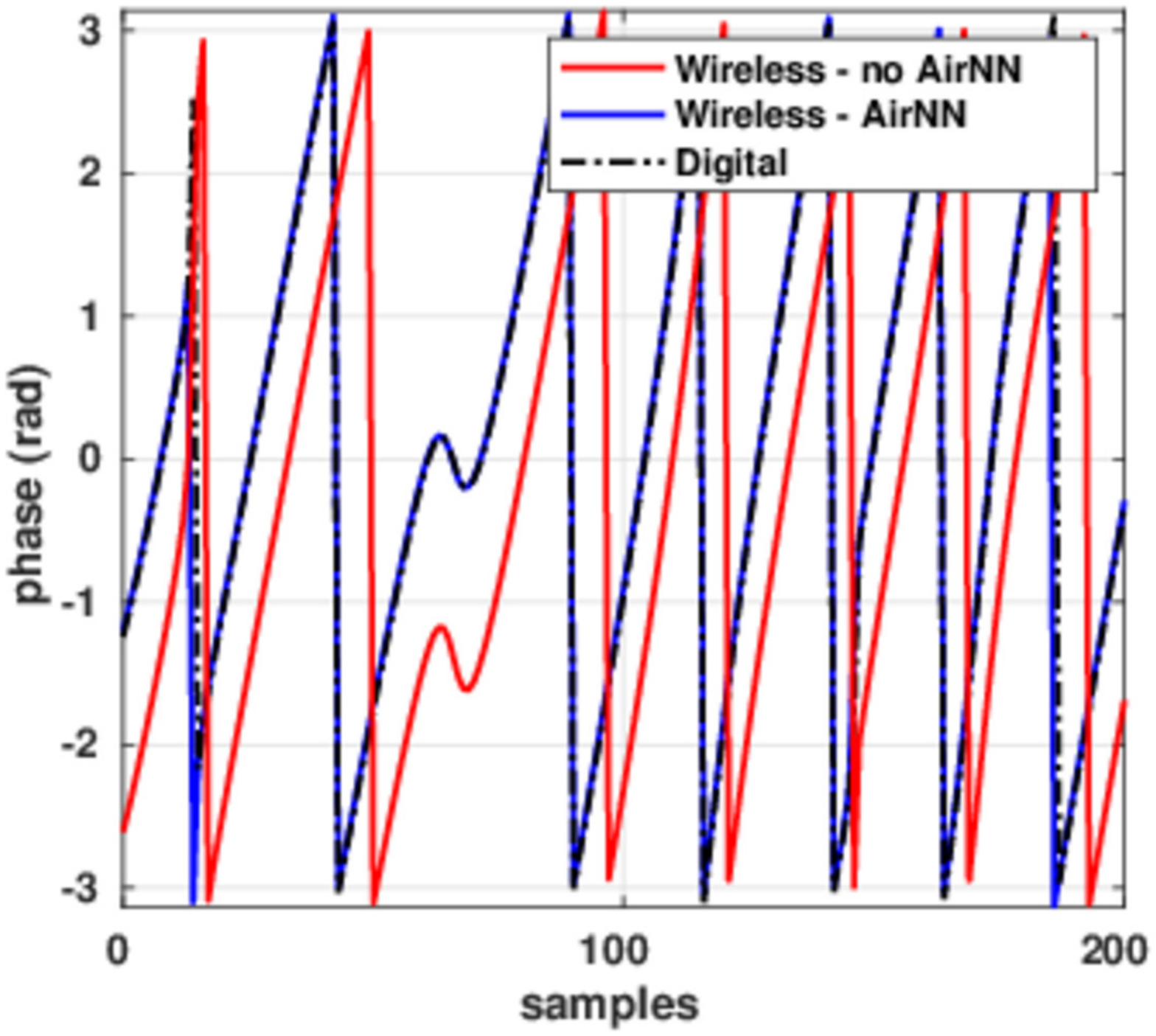}
  \vspace{-17mm}
  \caption{Phase}
  \label{fig:angleConv}
\end{subfigure}
\caption{\small  Comparison between over-the-air convolution with and without the use of AirNN RIS network. The former accurately realizes the desired convolutional filter with negligible error.}
\end{figure}

We first demonstrate the capability of AirNN to generate a signal that matches the expected output of an all-digital FIR filter in a CNN.
To this extent, AirNNOS first transmits unique GS and estimates the CIR trough cross-correlation and LS estimation at the R-Rx for all RIS configurations, as a one-time initial step.
 We then train our QM-CNN model and account for the filter weights that our RIS can provide. The relay then engineers an over-the-air convolution tailored to the desired FIR behavior.
At this stage, AirNNOS transmits a BPSK digitally modulated signal.
Finally, after the signal has traversed the three RIS, we store the received signal at the \texttt{R-Rx}, whose magnitude and phase are shown in blue colour in Figs. \ref{fig:magConv} and \ref{fig:angleConv}, respectively.
We compare the similarity of the received signal to that of an all-digital convolution, shown in black colour observing a Root Mean Square Error (RMSE) of $0.11$ and $0.6$ in magnitude and phase, respectively. When we store the received signal without controlling the RIS network (shown in red colour), the lack of temporal alignment and the RIS misconfiguration leads to a phase and magnitude mismatch with the all-digital convolution. This increases RMSE values to $ 0.46$ and  $2.58$ in phase and magnitude, respectively.

\subsection{AirNN for Modulation Classification}
\label{subsec:dataset}
Next, we demonstrate how the convolution performed in AirNN is accurate enough to replace its digital equivalent for a real-world problem of modulation classification.

\noindent$\bullet$\textbf{Dataset description:}
 We use the RADIOML 2018.01A dataset released in  \cite{o2018over}. This includes signals collected from over-the-air transmissions modulated with 24 different schemes, i.e., from BPSK to 256QAM, under variable link qualities or SNR levels that range from -10 to 30dB.
The data is organized in IQ sequences of 1024 I/Q samples, with 4096 sequences per modulation/SNR pair. Since this paper focuses on AirNN design (and not improving on best-performing architecture for the problem of modulation classification), we consider a smaller subset of the problem with four of the most common classes of BPSK QPSK, 16QAM, and 32QAM. 
We split this reduced dataset into non-overlapping portions for training (60\%), validation (20\%), and testing (20\%). 

\begin{figure}
    \centering
    \includegraphics[width=1\linewidth]{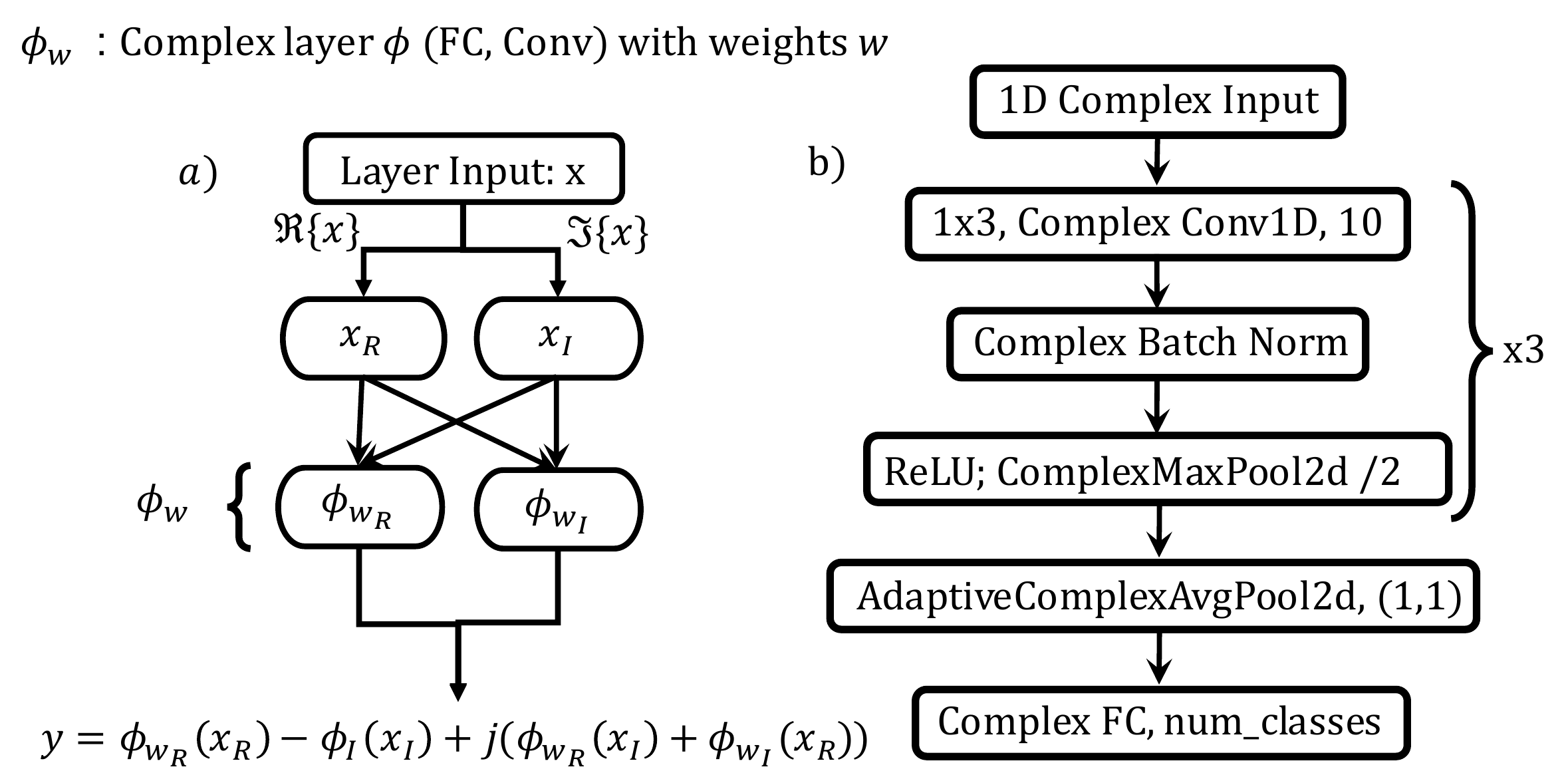}
    \caption{\small Complex-valued layer $(\phi_w)$ diagram (a) and neural network architecture with complex weights used for modulation classification (b).}
    \label{fig:baseline}
\end{figure}

\noindent$\bullet$\textbf{Architecture Description:} Our deep CNN model is composed of three sequential convolutional layers, i.e., a bundle of convolutional FIR filters followed by the pertinent batch normalization, activation (ReLu) and max pooling, an adaptive average pool layer and a single fully connected layer (Fig.~\ref{fig:baseline}b). We use PyTorch for implementation, with the number of filters as ten and learning rate 1$e^{-4}$.

\subsection{Impact of Quantization}
\begin{figure}[t!]
\begin{subfigure}{0.49\linewidth}
  \centering
  \includegraphics[trim=0 0 20 20, clip, width=\linewidth]{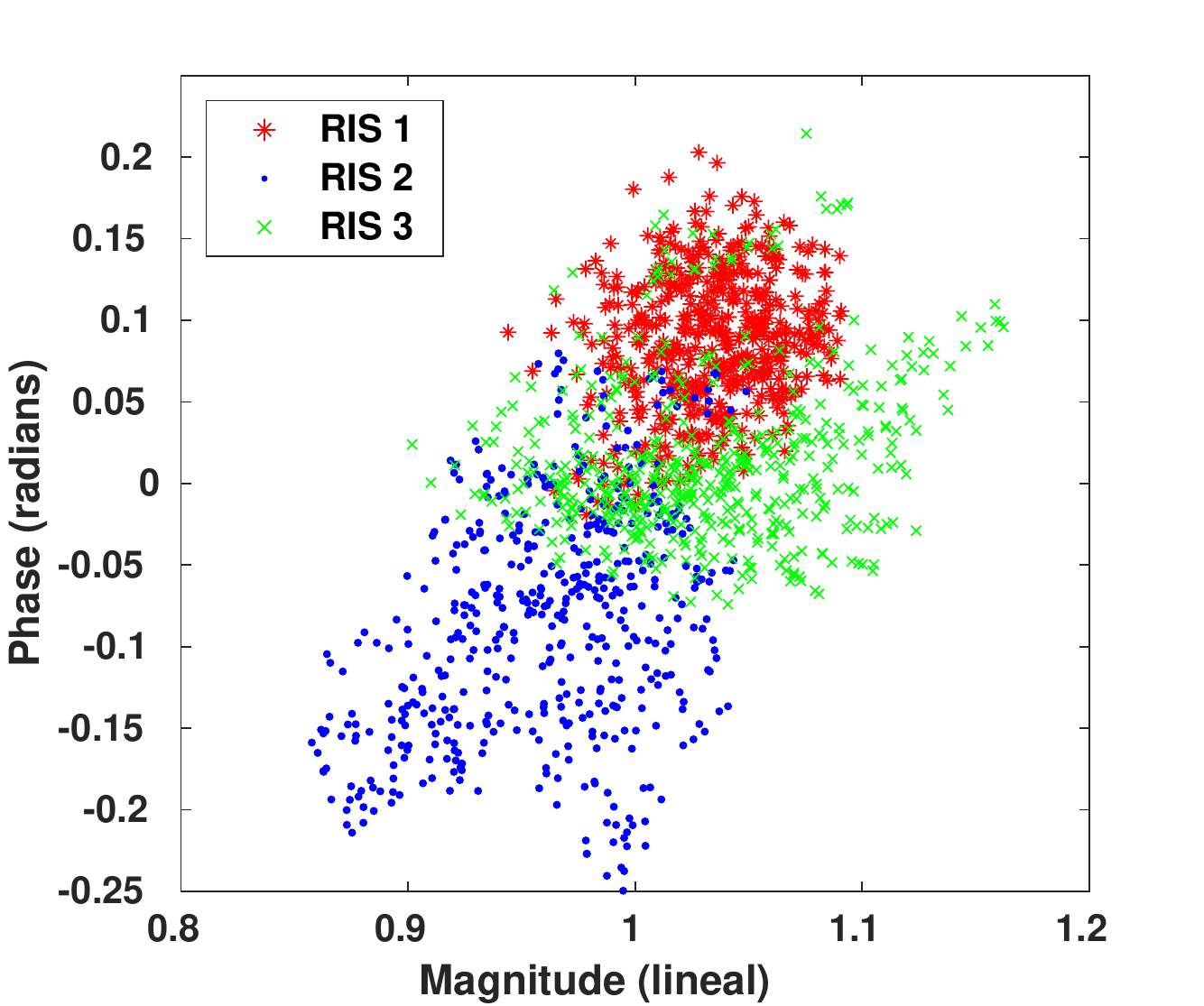}
  \caption{}
  \label{fig:mapping}
\end{subfigure}
\begin{subfigure}{0.49\linewidth}
  \centering
  \includegraphics[trim=0 0 20 20, clip, width=\linewidth]{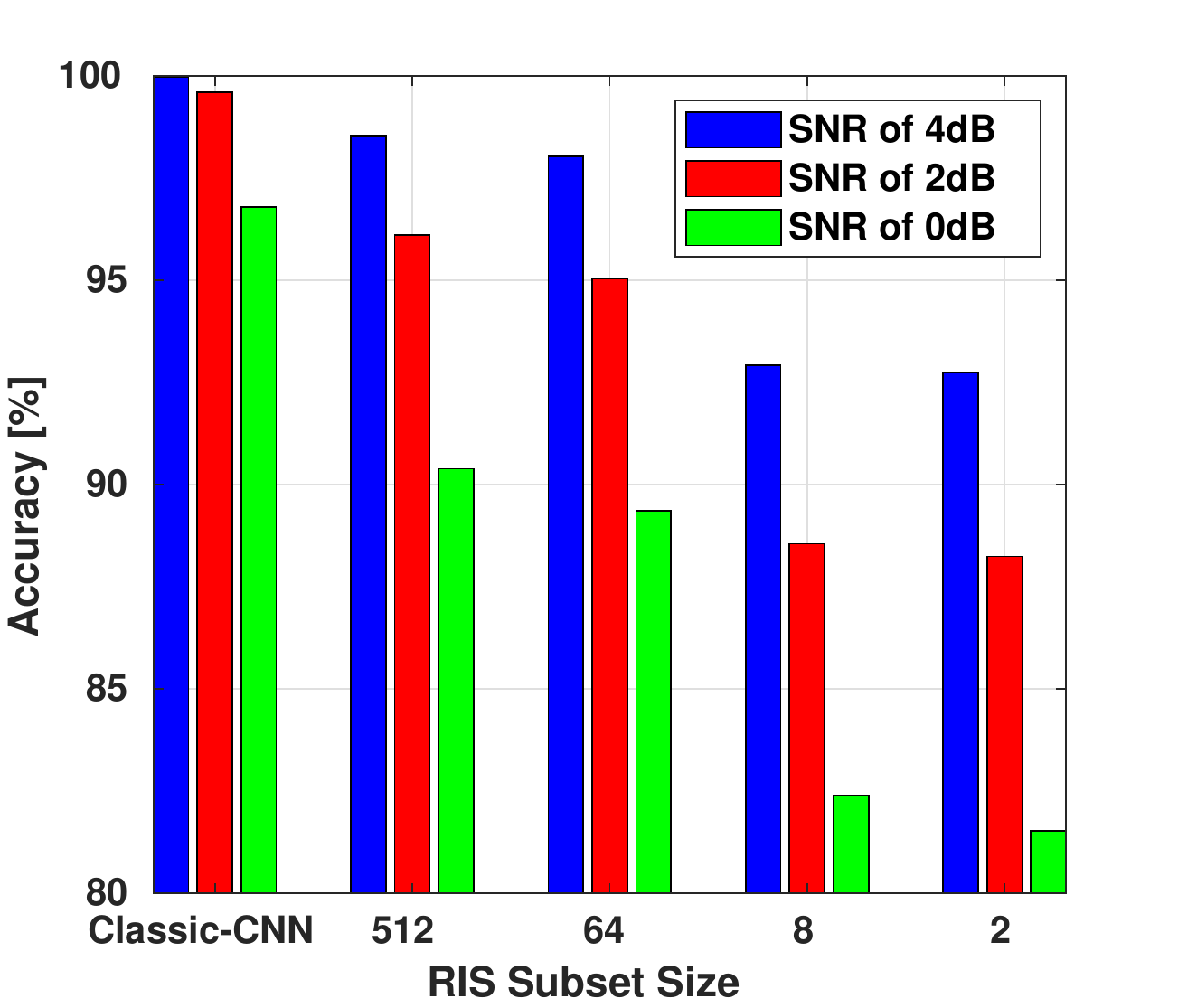}
  \caption{}
  \label{fig:impactQuanti}
\end{subfigure}
\vspace{-1mm}
\caption{\small (a) Realizable FIR filter taps (magnitude and phase) using RIS. Each RIS provides a total of 512 phase-amplitude selections, any of which can be used as a filter tap; (b) Effect of quantization on the accuracy in modulation classification for different SNR values.}
\end{figure}

Recall that we extract the set of feasible weights that our experimentally deployed RIS can realize, and use them to train our proposed QM-CNN (see Fig.~\ref{fig:baseline}).
For a tractable analysis, we consider the lower-end of the shifting range, i.e., $0\degree$ and $45\degree$, for each reflective element in all RIS.
This gives us a total of $2^9 = 512$ different phase shifts for the reflected signal.
We show the measured received CIR magnitude and phase at the \texttt{R-Rx} in Fig.~\ref{fig:mapping}, where each point is an average of over ten transmissions. Multiple works have explored different bit-level quantization-aware training, such as 4-bit \cite{zhou2016dorefa, choi2018pact, gong2019differentiable, cheng2019uL2Q}, 2-bit (ternary) \cite{li2016ternary, zhu2016trained, he2019simultaneously} or 1-bit (binary) \cite{courbariaux2016binarized, rastegari2016xnor, lin2017towards} while preserving non-quantized network performance. This is also the approach we use as the starting point of this work.

To assess how quantization impacts accuracy, we generate smaller sets of candidate weights by randomly selecting subsets of size $\{2,8,64\}$ from the global set of 512.
Here, $2$ represents the most restrictive (or quantized) case, implying that the entire QM-CNN is constructed with two possible weights for each convolutional filter tap.
Fig.~\ref{fig:impactQuanti} shows the average accuracy of QM-CNN when provided with various subset sizes for CIR weights and SNR values.
Note that the SNR captures the wireless link quality from the Tx to R-Tx (see Fig.~\ref{fig:systemArch}) and is provided by the dataset.
Reducing the set of realizable weights impacts the QM-CNN accuracy, which falls more than an 8\% for a quantization level below 64 and the lowest SNR evaluated of 0 dB. As the SNR increases up to 4 dB, the accuracy drops only a 2\% for quantization levels above 8, becoming only critical (7\%) for a quantization as low as eight levels and below.   

\subsection{Robustness to Noise}
\label{subsec:perfeval_robustness}
\begin{figure}[t!]
\begin{subfigure}{0.49\linewidth}
  \centering
  \includegraphics[trim=5 0 20 20, clip, width=\linewidth]{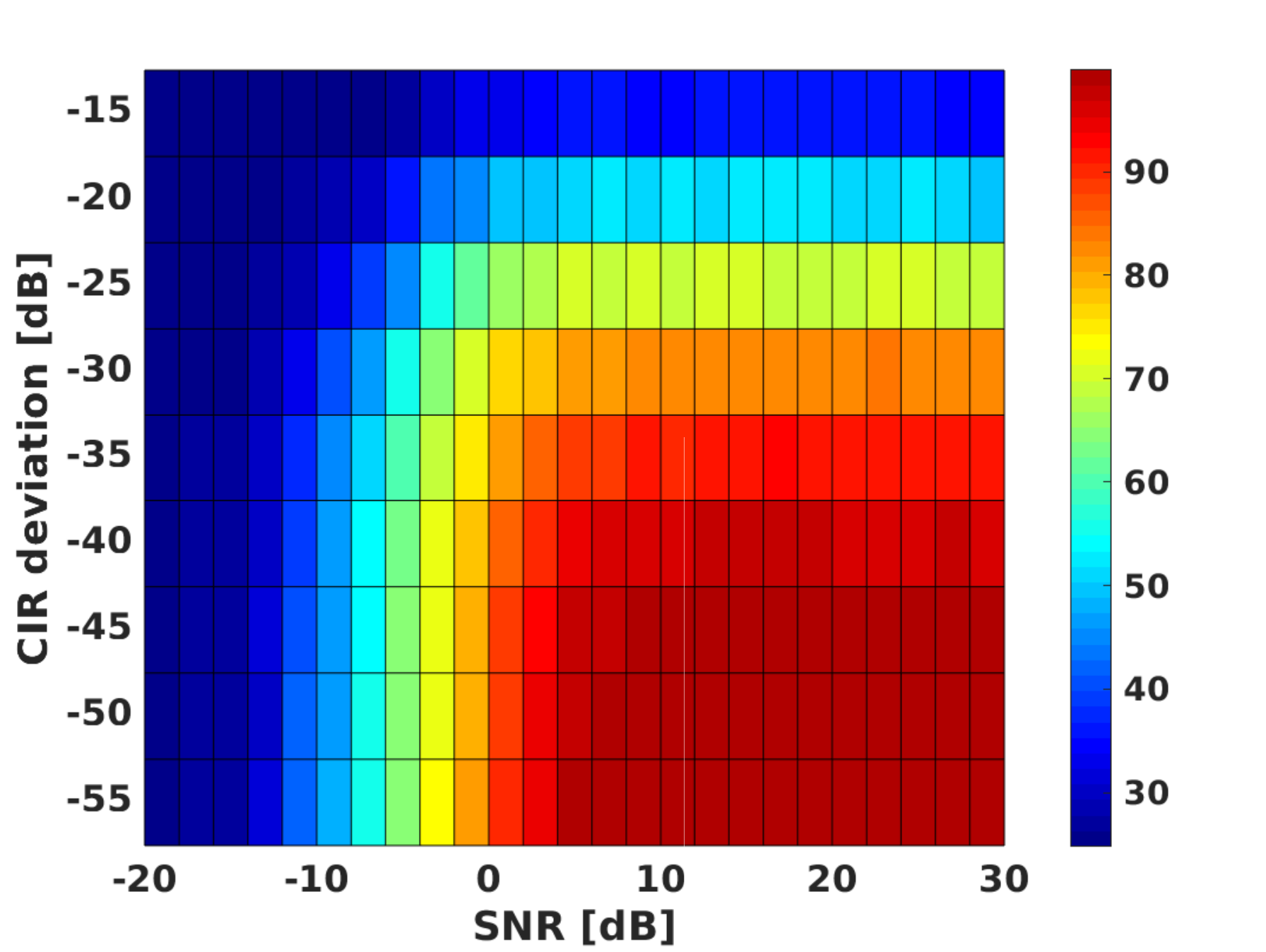}
  \caption{}
  \label{fig:QM_vs_noise}
\end{subfigure}
\begin{subfigure}{0.49\linewidth}
  \centering
  \includegraphics[trim=5 200 20 220, clip, width=\linewidth]{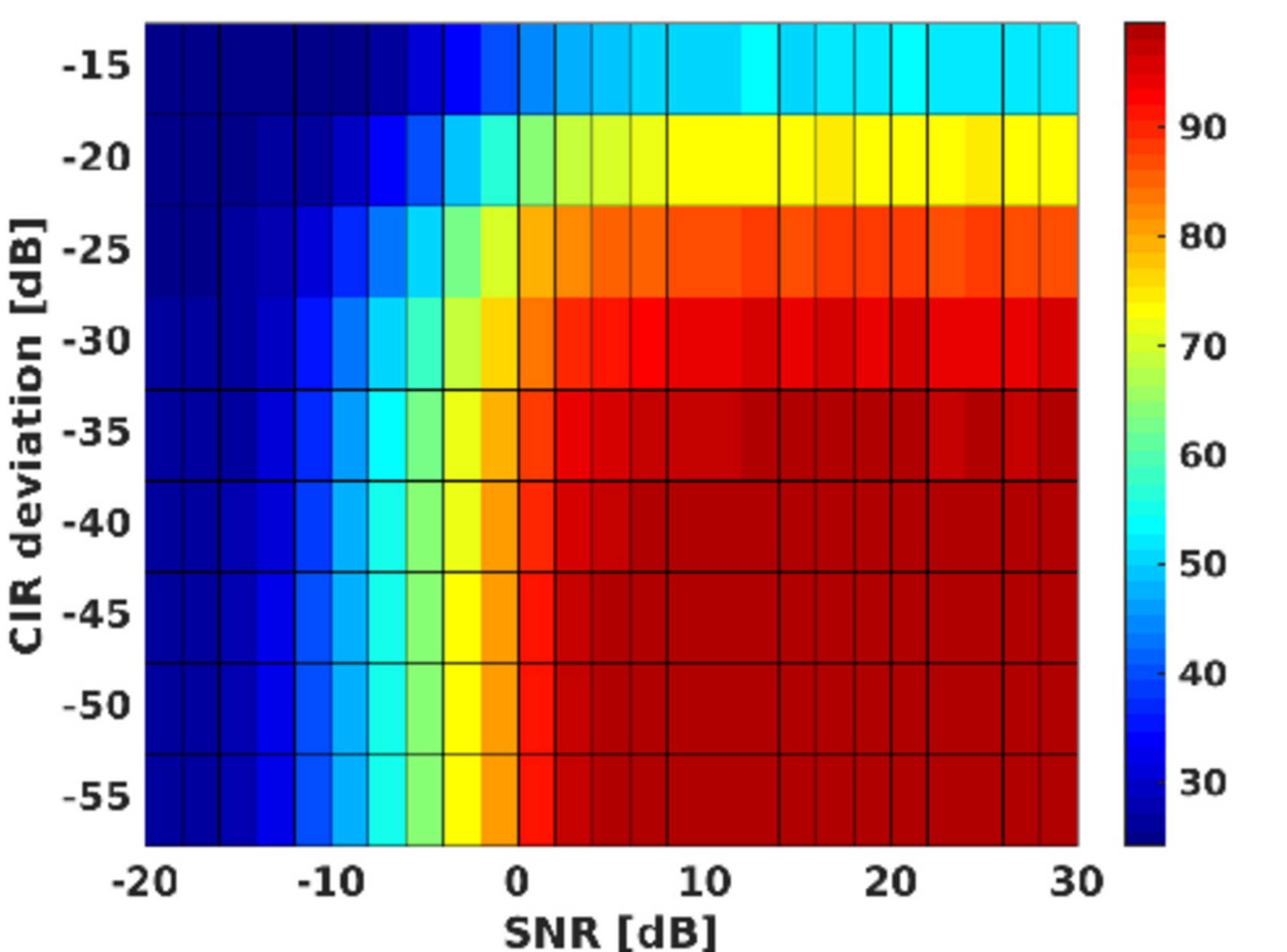}
  \caption{}
  \label{fig:RQM_vs_noise}
\end{subfigure}
\caption{\small Accuracy in modulation classification using (a) quantized model (QM-CNN) and (b) our quantized robust approach (RQM-CNN).}
\label{fig:accNoise}
\end{figure}

We validate the robustness of the proposed RQM-CNN approach, which aims to mitigate deviations in RIS-engineered CIR weights due to noise.
Fig.~\ref{fig:HCorrected} gives a visualization of such a deviation illustrating the worst-case deviation measured empirically on the complete secondary path: \texttt{R-Tx} to RIS to \texttt{R-Rx}, giving a CIR variance of -35 dB ($\sigma^2$ in Sec.~\ref{subsec:dataAug}). Once we profile a range of possible CIR deviations, we train our RQM-CNN under an AWGN distribution within this CIR variance bound, following the steps described in Sec.~\ref{subsec:dataAug}.
We test the RQM-CNN performance for SNR levels between -20 and 30 dB on the primary link given as: Tx to \texttt{R-Rx}, and CIR deviations between -55 and -15 dB over the  secondary link \texttt{R-Tx} to RIS to \texttt{R-Rx}. As opposed to this, simpler QM-CNN approach  does not account for such over-the-air impairments during training. Results shown in Fig.~\ref{fig:accNoise} reveal that QM-CNN provides good performance for higher SNR and CIR deviations, but does not provide accuracy
above 88\% for SNR levels below 4dB and CIR levels above -35 dB (see Fig.~\ref{fig:QM_vs_noise}). The RQM-CNN approach achieves an accuracy of up to 96\% in the same regimes (see Fig.~\ref{fig:RQM_vs_noise}).

\subsection{AirNN Performance}
In this section, we compare the accuracy between  the three all-digital CNN versions discussed so far: Classical-CNN and the  quantized versions QM-CNN and RQM-CNN, as a function of the SNR level of the Tx to \texttt{R-Tx} link. We also compare it with AirNN, using the experimentally derived convolution error on top of the RQM-CNN model. Here, QM-CNN and RQM-CNN are trained and tested as described in Sec.~\ref{subsec:perfeval_robustness}.
AirNN uses the same trained weights than RQM-CNN, but must operate within dynamic conditions that arise during testing. These modify the RIS-engineered FIR taps from the initial values acquired at the mapping stage (see Fig.~\ref{fig:HCorrected}), which we earlier characterized as AWGN (see Sec.~\ref{subsec:dataAug}).

\begin{figure}[t]
    \centering
    \includegraphics[trim=15 300 30 280, clip, width=.8\linewidth]{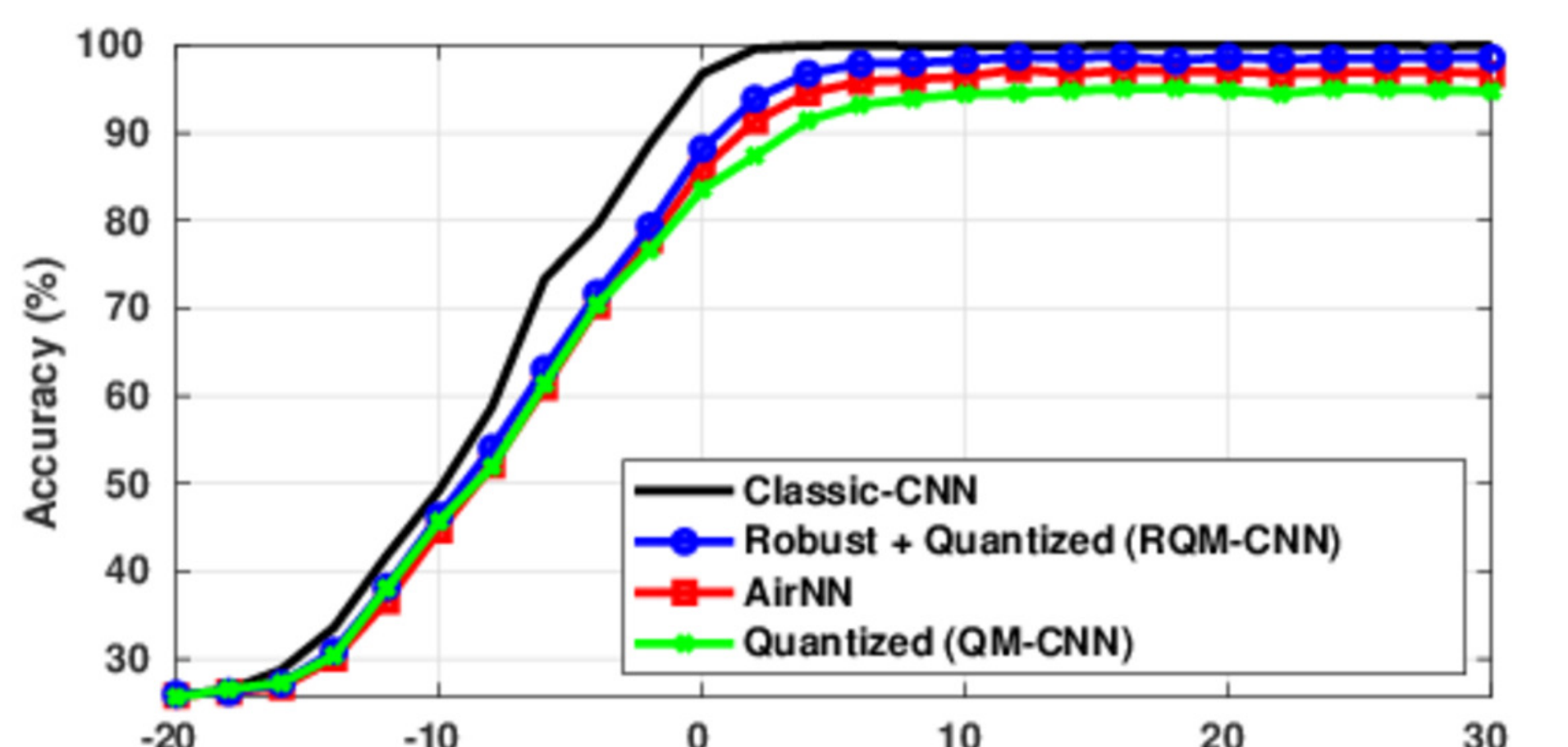}
    \vspace{4mm}
    \caption{\small Accuracy in modulation classification using all-digital convolution (Classic-CNN), quantized approaches QM-CNN and RQM-CNN, and our end-to-end AirNN system.}
    \centering
\label{fig:accSNR}
\end{figure}

We present the experimental results with AirNN in Fig.~\ref{fig:accSNR}, where the CIR inaccuracies are selected from a Gaussian PDF (see Sec.~\ref{sec:contrained_CNN}), ranging from -15dB to -50dB. In the figure, the Classical-CNN bounds the performance for any given SNR value.  We observe a similar accuracy reported from all four models for very low SNR, i.e., between -20dB and -5dB, which is extremely challenging for the classification task.
For higher SNR values, QM-CNN reports a lower maximum accuracy of 95\%, while the robust training in RQM-CNN raises the accuracy up to 98\%. AirNN closely follows the bound of the software-based RQM-CNN, with a drop in accuracy of only 2\%, and an overall drop w.r.t. Classic-CNN of 3.2\% for the SNR range of [6, 30] dB.

%% file: sections/conclusions.tex
\section{Limitations and Future Opportunities}
In this section we identify some limitations of our approach and identify candidate solutions for open challenges that will speed up practical deployment of AirNN. 

\noindent $\bullet$ AirNN uses \textit{N} directional antennas to compute an over-the-air convolution. This specific implementation requires tight synchronization between transmitters and needs to scale in terms of cost and complexity with the FIR filter size \textit{(N)}. This can be potentially addressed using massive multi-user MIMO that can construct many simultaneous beams and ubiquitous deployment of RIS in the ambient environment. Although our validation is limited in scale, there is potential for generalization in a more resource-rich network.

\noindent $\bullet$    The existence of grating lobes in off-the-shelf directional antennas that we use and the limited reflected power from our custom designed RIS constrain the relative location and separation distance between transmitters, receiver and RIS. Thus, they have to be carefully deployed to avoid interference and receive sufficient power. This situation can be mitigated with highly precise beams, improved RIS design with minimal losses, and scenarios where higher power transmission become possible, e.g., in outdoor environments. 
    
\noindent $\bullet$ Although AirNNOS is designed to tackle channel changes over time, it assumes that transmitter, receiver and RIS locations are fixed for both training and testing. Handling mobility requires interdisciplinary research in rapid wireless CIR estimation and lifelong-learning methods for RF pioneered by the ML community. 
     
\noindent $\bullet$ AirNN only emulates  the convolutional operation. It remains an open challenge to extend these ideas towards a complete CNN, all performed within the RF domain, which includes multi-layer convolutions and nonlinear activation functions. 
   
\noindent $\bullet$ Performing a convolution over-the-air involves higher power consumption and latency compared to its digital equivalent running on a GPU or FPGA. Our approach is best suited for scenarios where signals must be transmitted over-the-air and AirNN merely provides an \textit{add-on} functionality. Thus, supporting network infrastructure like multiple synchronized antennas and RIS should not be solely present to realize AirNN. 

Despite the above limitations, AirNN points towards an exciting computational domain involving RF signals. We will open-source design files for the RIS, code for AirNNOS and RIS simulation tool to equip the community with essential tools for future systems-focused work that can lead to full-fledged over-the-air CNNs.

  \section{Conclusions}
    \label{sec:Conclusions}
    We have demonstrated the feasibility of engineering convolution operations over-the-air through a programmable RIS network that is precisely equivalent to its digital counterpart. We report an RMSE of $0.11$ and $0.6$ in magnitude and phase, respectively, in the output of the convolution between the analog and digital versions. Furthermore, we have shown how this operation, when included within the processing steps of a trained CNN is accurate enough to run inference on signal analysis tasks such as modulation classification. AirNN average testing accuracy is within $3.2$\% of the classical digital version under medium-to-high SNR conditions.

%% file: main.bbl
\begin{thebibliography}{10}

\bibitem{hughes2019wave}
T.~W. Hughes, I.~A. Williamson, M.~Minkov, and S.~Fan, ``Wave Physics as an Analog Recurrent Neural Network,'' {\em Science advances}, vol.~5, no.~12,
  p.~eaay6946, 2019.

\bibitem{sui2020review}
X.~Sui, Q.~Wu, J.~Liu, Q.~Chen, and G.~Gu, ``A Review of Optical Neural Networks,'' {\em IEEE Access}, vol.~8, pp.~70773--70783, 2020.

\bibitem{elbir2020survey}
A.~M. Elbir and K.~V. Mishra, ``A Survey of Deep Learning Architectures for Intelligent Reflecting Surfaces,'' {\em arXiv preprint arXiv:2009.02540},
  2020.

\bibitem{yu2020optimizing}
D.~Yu, S.-H. Park, O.~Simeone, and S.~S. Shitz, ``Optimizing Over-the-Air Computation in IRS-Aided C-RAN Systems,'' in {\em 2020 IEEE 21st
  International Workshop on Signal Processing Advances in Wireless
  Communications (SPAWC)}, pp.~1--5, IEEE, 2020.

\bibitem{zhao2017waveforms}
B.~Zhao, S.~Xiao, H.~Lu, and J.~Liu, ``Waveforms classification based on
  convolutional neural networks,'' in {\em 2017 IEEE 2nd Advanced Information
  Technology, Electronic and Automation Control Conference (IAEAC)},
  pp.~162--165, IEEE, 2017.

\bibitem{cai2019modulation}
J.~Cai, C.~Li, and H.~Zhang, ``Modulation Recognition of Radar Signal Based on an Improved CNN Model,'' in {\em 2019 IEEE 7th International Conference on
  Computer Science and Network Technology (ICCSNT)}, pp.~293--297, IEEE, 2019.

\bibitem{restuccia2019deepradioid}
F.~Restuccia, S.~D'Oro, A.~Al-Shawabka, M.~Belgiovine, L.~Angioloni,
  S.~Ioannidis, K.~Chowdhury, and T.~Melodia, ``DeepRadioID: Real-Time Channel-Resilient Optimization of Deep Learning-based Radio Fingerprinting Algorithms,'' in {\em Proceedings of the Twentieth ACM International
  Symposium on Mobile Ad Hoc Networking and Computing}, pp.~51--60, 2019.

\bibitem{dunna2020scattermimo}
M.~Dunna, C.~Zhang, D.~Sievenpiper, and D.~Bharadia, ``ScatterMIMO: Enabling Virtual MIMO with Smart Surfaces,'' in {\em Proceedings of the 26th Annual
  International Conference on Mobile Computing and Networking}, pp.~1--14,
  2020.

\bibitem{boser1991analog}
B.~E. Boser, E.~Sackinger, J.~Bromley, Y.~Le~Cun, and L.~D. Jackel, ``An analog
  neural network processor with programmable topology,'' {\em IEEE Journal of
  Solid-State Circuits}, vol.~26, no.~12, pp.~2017--2025, 1991.

\bibitem{misra2010artificial}
J.~Misra and I.~Saha, ``Artificial neural networks in hardware: A survey of two
  decades of progress,'' {\em Neurocomputing}, vol.~74, no.~1-3, pp.~239--255,
  2010.

\bibitem{kendall2020training}
J.~Kendall, R.~Pantone, K.~Manickavasagam, Y.~Bengio, and B.~Scellier,
  ``Training End-to-End Analog Neural Networks with Equilibrium Propagation,''
  {\em arXiv preprint arXiv:2006.01981}, 2020.

\bibitem{li2018efficient}
C.~Li, D.~Belkin, Y.~Li, P.~Yan, M.~Hu, N.~Ge, H.~Jiang, E.~Montgomery, P.~Lin,
  Z.~Wang, {\em et~al.}, ``Efficient and self-adaptive in-situ learning in
  multilayer memristor neural networks,'' {\em Nature communications}, vol.~9,
  no.~1, pp.~1--8, 2018.

\bibitem{james2021analog}
A.~P. James and L.~O. Chua, ``Analog Neural Computing with Super-resolution Memristor Crossbars,'' {\em IEEE Transactions on Circuits and Systems I:
  Regular Papers}, 2021.

\bibitem{du2018analog}
Y.~Du, L.~Du, X.~Gu, J.~Du, X.~S. Wang, B.~Hu, M.~Jiang, X.~Chen, S.~S. Iyer,
  and M.-C.~F. Chang, ``An Analog Neural Network Computing Engine using CMOS-Compatible Charge-Trap-Transistor (CTT),'' {\em IEEE Transactions on
  Computer-Aided Design of Integrated Circuits and Systems}, vol.~38, no.~10,
  pp.~1811--1819, 2018.

\bibitem{cauwenberghs1996analog}
G.~Cauwenberghs, ``An Analog VLSI Recurrent Neural Network Learning a
  Continuous-Time Trajectory,'' {\em IEEE Transactions on Neural Networks},
  vol.~7, no.~2, pp.~346--361, 1996.

\bibitem{umuroglu2020logicnets}
Y.~Umuroglu, Y.~Akhauri, N.~J. Fraser, and M.~Blott, ``LogicNets: Co-Designed Neural Networks and Circuits for Extreme-Throughput Applications,'' in {\em
  2020 30th International Conference on Field-Programmable Logic and
  Applications (FPL)}, pp.~291--297, IEEE, 2020.

\bibitem{strukov2019building}
D.~Strukov, G.~Indiveri, J.~Grollier, and S.~Fusi, ``Building brain-inspired
  computing,'' {\em Nature Communications}, no.~10, pp.~4838--2019, 2019.

\bibitem{wen2019reduced}
D.~Wen, G.~Zhu, and K.~Huang, ``Reduced-Dimension Design of MIMO Over-the-Air Computing for Data Aggregation in Clustered IoT Networks,'' in {\em 2019 IEEE Global
  Communications Conference (GLOBECOM)}, pp.~1--6, IEEE, 2019.

\bibitem{wang2020wirelessly}
Z.~Wang, Y.~Shi, and Y.~Zhou, ``Wirelessly Powered Data Aggregation via Intelligent Reflecting Surface Assisted Over-the-Air Computation,'' in {\em
  2020 IEEE 91st Vehicular Technology Conference (VTC2020-Spring)}, pp.~1--5,
  IEEE, 2020.

\bibitem{arun2020rfocus}
V.~Arun and H.~Balakrishnan, ``{RFocus: Beamforming Using Thousands
of Passive Antennas},'' in {\em 17th $\{$USENIX$\}$ Symposium on Networked Systems
  Design and Implementation ($\{$NSDI$\}$ 20)}, pp.~1047--1061, 2020.

\bibitem{liu2021joint}
H.~Liu, X.~Yuan, and Y.-J.~A. Zhang, ``Joint Communication-Learning Design for RIS-Assisted Federated Learning,'' in {\em 2021 IEEE International Conference
  on Communications Workshops (ICC Workshops)}, pp.~1--6, IEEE, 2021.

\bibitem{ni2020intelligent}
W.~Ni, Y.~Liu, and H.~Tian, ``Intelligent Reflecting Surfaces Enhanced Federated Learning,'' in {\em 2020 IEEE Globecom Workshops (GC Wkshps},
  pp.~1--6, IEEE, 2020.

\bibitem{wang2021federated}
Z.~Wang, J.~Qiu, Y.~Zhou, Y.~Shi, L.~Fu, W.~Chen, and K.~B. Letaief,
  ``Federated Learning via Intelligent Reflecting Surface,'' {\em IEEE
  Transactions on Wireless Communications}, 2021.

\bibitem{jiang2019over}
T.~Jiang and Y.~Shi, ``Over-the-Air Computation via Intelligent Reflecting Surfaces,'' in {\em 2019 IEEE Global Communications Conference (GLOBECOM)},
  pp.~1--6, IEEE, 2019.

\bibitem{ni2021over}
W.~Ni, Y.~Liu, Z.~Yang, and H.~Tian, ``Over-the-Air Federated Learning and Non-Orthogonal Multiple Access Unified by Reconfigurable Intelligent Surface,'' in {\em IEEE INFOCOM 2021-IEEE Conference on Computer
  Communications Workshops (INFOCOM WKSHPS)}, pp.~1--6, IEEE, 2021.

\bibitem{lecun2015deep}
Y.~LeCun, Y.~Bengio, and G.~Hinton, ``Deep learning,'' {\em nature}, vol.~521,
  no.~7553, pp.~436--444, 2015.

\bibitem{li2016performance}
X.~Li, G.~Zhang, H.~H. Huang, Z.~Wang, and W.~Zheng, ``Performance Analysis of GPU-Based Convolutional Neural Networks,'' in {\em 2016 45th International
  conference on parallel processing (ICPP)}, pp.~67--76, IEEE, 2016.

\bibitem{trabelsi2018deep}
C.~Trabelsi, O.~Bilaniuk, Y.~Zhang, D.~Serdyuk, S.~Subramanian, J.~F. Santos,
  S.~Mehri, N.~Rostamzadeh, Y.~Bengio, and C.~J. Pal, ``Deep Complex
  Networks,'' in {\em International Conference on Learning Representations},
  2018.

\bibitem{DBLP:journals/corr/BengioLC13}
Y.~Bengio, N.~L{\'e}onard, and A.~Courville, ``Estimating or Propagating Gradients Through Stochastic Neurons for Conditional Computation,'' {\em
  arXiv preprint arXiv:1308.3432}, 2013.

\bibitem{DBLP:journals/corr/abs-1903-05662}
P.~Yin, J.~Lyu, S.~Zhang, S.~Osher, Y.~Qi, and J.~Xin, ``Understanding Straight-Through Estimator in Training Activation Quantized Neural Nets,'' in
  {\em International Conference on Learning Representations (ICLR)}, 2018.

\bibitem{chang2020mix}
S.-E. Chang, Y.~Li, M.~Sun, R.~Shi, H.~K.-H. So, X.~Qian, Y.~Wang, and X.~Lin,
  ``Mix and Match: A Novel FPGA-Centric Deep Neural Network Quantization Framework,'' {\em arXiv preprint arXiv:2012.04240}, 2020.

\bibitem{Gaussian}
X.~Zhang, {\em Gaussian Distribution}, pp.~425--428.
\newblock Boston, MA: Springer US, 2010.

\bibitem{ozdogan2019intelligent}
{\"O}.~{\"O}zdogan, E.~Bj{\"o}rnson, and E.~G. Larsson, ``{Intelligent Reflecting Surfaces: Physics, Propagation, and Pathloss Modeling},'' {\em
  IEEE Wireless Communications Letters}, vol.~9, no.~5, pp.~581--585, 2019.

\bibitem{balanis2016antenna}
C.~A. Balanis, {\em Antenna Theory: Analysis and Design}.
\newblock John wiley \& sons, 2016.

\bibitem{rfswitch}
A.~Devices, ``HMC7992: Non-Reflective, Silicon SP4T Switch, 0.1 GHz to 6.0
  GHz.''

\bibitem{Xinyu2011AnalysisSystem}
Z.~Xinyu, ``{Analysis of M-sequence and Gold-sequence in CDMA system},'' {\em
  IEEE International Conference on Communication Software and Networks
  (ICCSN)}, no.~1, pp.~466--468, 2011.

\bibitem{o2018over}
T.~J. O’Shea, T.~Roy, and T.~C. Clancy, ``{Over the Air Deep Learning
Based Radio Signal Classification},'' {\em IEEE Journal of Selected Topics in
  Signal Processing}, vol.~12, no.~1, pp.~168--179, 2018.

\bibitem{zhou2016dorefa}
S.~Zhou, Y.~Wu, Z.~Ni, X.~Zhou, H.~Wen, and Y.~Zou, ``DoReFa-Net: Training Low Bitwidth Convolutional Neural Networks with Low Bitwidth Gradients,'' {\em
  arXiv preprint arXiv:1606.06160}, 2016.

\bibitem{choi2018pact}
J.~Choi, Z.~Wang, S.~Venkataramani, P.~I.-J. Chuang, V.~Srinivasan, and
  K.~Gopalakrishnan, ``PACT: Parameterized Clipping Activation for Quantized Neural Networks,'' {\em arXiv preprint arXiv:1805.06085}, 2018.

\bibitem{gong2019differentiable}
R.~Gong, X.~Liu, S.~Jiang, T.~Li, P.~Hu, J.~Lin, F.~Yu, and J.~Yan,
  ``Differentiable Soft Quantization:
Bridging Full-Precision and Low-Bit Neural Networks,'' in {\em Proceedings of the IEEE International Conference
  on Computer Vision (ICCV)}, pp.~4852--4861, 2019.

\bibitem{cheng2019uL2Q}
G.~Cheng, L.~Ye, L.~Tao, Z.~Xiaofan, H.~Cong, C.~Deming, and C.~Yao,
  ``$\mu$L2Q: An Ultra-Low Loss Quantization Method for DNN Compression,'' {\em The 2019
  International Joint Conference on Neural Networks (IJCNN)}, 2019.

\bibitem{li2016ternary}
F.~Li, B.~Zhang, and B.~Liu, ``Ternary Weight Networks,'' {\em arXiv preprint
  arXiv:1605.04711}, 2016.

\bibitem{zhu2016trained}
C.~Zhu, S.~Han, H.~Mao, and W.~J. Dally, ``Trained Ternary Quantization,'' in
  {\em International Conference on Learning Representations (ICLR)}, 2017.

\bibitem{he2019simultaneously}
Z.~He and D.~Fan, ``Simultaneously Optimizing Weight and Quantizer of Ternary Neural Network using Truncated Gaussian Approximation,'' in {\em Proceedings
  of the IEEE Conference on Computer Vision and Pattern Recognition (CVPR)},
  pp.~11438--11446, 2019.

\bibitem{courbariaux2016binarized}
M.~Courbariaux, I.~Hubara, D.~Soudry, R.~El-Yaniv, and Y.~Bengio, ``Binarized Neural Networks: Training Deep Neural Networks with Weights and Activations Constrained to +1 or -1,'' {\em arXiv preprint arXiv:1602.02830}, 2016.

\bibitem{rastegari2016xnor}
M.~Rastegari, V.~Ordonez, J.~Redmon, and A.~Farhadi, ``XNOR-Net: ImageNet Classification Using Binary Convolutional Neural Networks,'' in {\em European
  conference on computer vision (ECCV)}, pp.~525--542, Springer, 2016.

\bibitem{lin2017towards}
X.~Lin, C.~Zhao, and W.~Pan, ``Towards Accurate Binary Convolutional Neural Network,'' in {\em Advances in Neural Information Processing Systems
  (NeurIPS)}, pp.~345--353, 2017.

\end{thebibliography}
